\documentclass[]{aa}
\usepackage{natbib}
\bibpunct{(}{)}{;}{a}{}{,} 
\usepackage{graphicx}
\usepackage{txfonts}
\usepackage[]{hyperref}
\usepackage{lineno}

\begin{document}

   \title{Applying a temporal systematics model to vector Apodizing Phase Plate coronagraphic data: TRAP4vAPP}


   \author{Pengyu Liu 
          \inst{1,2,3}
          \and 
          Alexander J. Bohn\inst{1}
          \and 
          David S. Doelman\inst{1}
          \and 
          Ben J. Sutlieff\inst{2,3,4,1}
          \and
          Matthias Samland\inst{5}
          \and 
          Matthew A. Kenworthy\inst{1}
          \and 
          Frans Snik\inst{1}
          \and
          Jayne L. Birkby\inst{6}
          \and
          Beth A. Biller\inst{2,3}
          \and
          Jared R. Males\inst{7}
          \and
          Katie M. Morzinski\inst{7}
          \and
          Laird M. Close\inst{7}
          \and
          Gilles P. P. L. Otten\inst{1,8}
          }

   \institute{Leiden Observatory, Leiden University, PO Box 9513, 2300 RA Leiden, The Netherlands\\
              \email{pengyu.liu@ed.ac.uk}
        \and
            SUPA, Institute for Astronomy, University of Edinburgh, Royal Observatory, Blackford Hill, Edinburgh EH9 3HJ, UK
        \and
            Centre for Exoplanet Science, University of Edinburgh, Edinburgh, UK
         \and
             Anton Pannekoek Institute for Astronomy, University of Amsterdam, Science Park 904, 1098 XH Amsterdam, The Netherlands
        \and
            Max-Planck-Institut für Astronomie, Königstuhl 17, 69117 Heidelberg, Germany
        \and 
            Astrophysics, University of Oxford, Denys Wilkinson Building, Keble Road, Oxford OX1 3RH, UK
        \and
            Steward Observatory, University of Arizona, 933 N. Cherry Ave., Tucson, AZ 85721, USA
        \and
            Academia Sinica, Institute of Astronomy and Astrophysics, 11F Astronomy-Mathematics Building, NTU/AS campus, No. 1, Section 4, Roosevelt Rd., Taipei 10617, Taiwan
             }

   \date{Received 31 October 2022 / Accepted 24 April 2023}

 
  \abstract
   {
   The vector Apodizing Phase Plate (vAPP) is a pupil plane coronagraph that suppresses starlight by forming a dark hole in its point spread function (PSF).
   The unconventional and non-axisymmetrical PSF arising from the phase modification applied by this coronagraph presents a special challenge to post-processing techniques.
   }
   {We aim to implement a recently developed post-processing algorithm, temporal reference analysis of planets (TRAP) on vAPP coronagraphic data. The property of TRAP that uses non-local training pixels, combined with the unconventional PSF of vAPP, allows for more flexibility than previous spatial algorithms in selecting reference pixels to model systematic noise.
  }
   {Datasets from two types of vAPPs are analysed: a double grating-vAPP (dgvAPP360) that produces a single symmetric PSF and a grating-vAPP (gvAPP180) that produces two D-shaped PSFs. We explore how to choose reference pixels to build temporal systematic noise models in TRAP for them. We then compare the performance of TRAP with previously implemented algorithms that produced the best signal-to-noise ratio (S/N) in companion detections in these datasets.}
   {We find that the systematic noise between the two D-shaped PSFs is not as temporally associated as expected. Conversely, there is still a significant number of systematic noise sources that are shared by the dark hole and the bright side in the same PSF. We should choose reference pixels from the same PSF when reducing the dgvAPP360 dataset or the gvAPP180 dataset with TRAP. In these datasets, TRAP achieves results consistent with previous best detections, with an improved S/N for the gvAPP180 dataset.
   }
   {}


   \titlerunning{TRAP4vAPP}
   \maketitle
%

\section{Introduction}
\label{sec:Introduction}
Direct imaging is a powerful technique for characterising the properties of exoplanets, such as their effective temperature, atmospheric composition, orbital motion, and top-of-atmosphere inhomogeneity \citep[e.g.][]{Skemer2014, Macintosh2015, Rajan2017, Samland2017, Wang2018, Biller2021}. However, direct imaging of exoplanets is challenging due to the high contrast and small angular separation between the planet and its host star. Overcoming this challenge requires the development of world-class instrumentation, carefully designed observation strategies, and advanced post-processing algorithms.
Dozens of exoplanets and sub-stellar companions have been imaged so far, such as the famous multi-planetary system, HR~8799~bcde \citep{Marois2008,Marois2010}, $\beta$~Pictoris~b \citep{Lagrange2010}, and 51~Eridani~b \citep{Macintosh2015}. Surveys such as the SpHere INfrared Exoplanets (SHINE) project \citep{Chauvin2017,Desidera2021,Langlois2021,Vigan2021}, Gemini Planet Imager Exoplanet Survey \citep[GPIES,][]{Nielsen2013,Nielsen2019}, B-star Exoplanet Abundance STudy \citep[BEAST,][]{Janson2019}, Young Suns Exoplanet Survey \citep[YSES,][]{Bohn2020a,Bohn2020b,Bohn2021}, and Code for Orbital Parametrization of Astrometrically Inferred New Systems (COPAINS) pilot survey \citep{Bonavita2022} are shaping our general understanding of the giant exoplanet and host star demographics.

Coronagraphs are important components of the instrumentation used in direct imaging. They suppress a significant amount of stellar flux while maximising the throughput of planet flux with a tradeoff between contrast, angular resolution, and working angles. The vector Apodizing Phase Plate (vAPP), the successor to the Apodizing Phase Plate (APP) \citep{Kenworthy2007}, is a pupil plane coronagraph that uses liquid crystals and direct writing techniques to modify the phase of incoming light. This produces a dark zone in the coronagraphic point spread function (PSF) where high-contrast companions can be detected \citep{Snik2012, Otten2014}. Compared with focal plane coronagraphs, vAPPs are more stable to tip-tilt instability. Unlike the PSF of APP, the PSF of vAPP has a 360-degree coverage around the central star. Depending on its design, there are several types of vAPPs that produce different shapes of dark zones and PSFs \citep{Doelman2021}. Two common types of vAPPs currently installed on telescopes are the grating-vAPP (gvAPP180), which produces two $180^{\circ}$ or D-shaped dark holes, and the double grating-vAPP (dgvAPP360), which produces a $360^{\circ}$ circular dark hole.

Although vAPP coronagraphs have been developed for about ten years, a consensus on the optimal post-processing strategy for vAPP data has yet to be reached. In particular, the two complementary coronagraphic PSFs of gvAPP180 add an additional challenge to the data reduction. Several techniques before this work have been applied to the vAPP data. For example, \cite{Otten2017} exploited the symmetry of the two PSFs of gvAPP180. They reduced the first on-sky dataset of the gvAPP180 on MagAO/Clio2 by rotating, scaling, and subtracting one PSF from its complementary PSF. \cite{Sutlieff2021} applied three techniques to their \object{HR~2562} dataset also obtained by the same gvAPP180. The first one is to join the two dark holes from the two complementary PSFs and implement classical angular differential imaging (cADI). The second technique, ADI + principal component analysis (PCA), also operates on the joined dark holes but uses PCA to model and subtract speckles before derotation. The last technique is flipped differential imaging (FDI + PCA) which uses the symmetry of the two complementary PSFs. It rotates the PSFs of one side by 180 degrees, implements PCA on them to build a reference PSF for the opposite PSFs, and then subtracts it. This technique is similar to the method used by \cite{Otten2017}. They find that with these three methods, cADI produces the strongest detection of the companion.
\cite{Otten2017} and \cite{Sutlieff2021} find that the symmetry between the two complementary PSFs produced by gvAPP180 is not as good as expected.
\cite{Wagner2020} used dgvAPP360 on the Large Binocular Telescope (LBT) to observe the protoplanetary disk around PDS~201 and reduced the dataset with the package Karhunen–Loève Image Projection (KLIP) \citep{Soummer2012, Apai2016}. \cite{Doelman2022} designed an ADI + PCA based technique to reduce the $L$ band integral field spectroscopy (IFS) dataset of \object{HR~8799} obtained from the dgvAPP360 mounted on LBT.

The above algorithms are all spatial algorithms. The recent emergence of a few time domain algorithms, such as wavelet-based temporal suppression for exoplanet detection \citep{Bonse2018}, exoplanet detection based on PAtch COvariance (PACO) \citep{Flasseur2018}, temporal reference analysis of planets (TRAP) \citep{Samland2021}, half-sibling regression on exoplanet imaging \citep{Gebhard2022}, speckle space-time covariance in high-contrast imaging \citep{Lewis2023}, and PCA-Temporal \citep{Long2023}, has piqued interest in exploiting the temporal correlation of noise in high-contrast imaging. 
The TRAP algorithm is an inverse technique that reconstructs the planet signal and systematic noise simultaneously in a linear regression model. Unlike spatial algorithms that use local pixels from other images to build a reference spatial model, TRAP uses non-local but causally related pixels to build a reference temporal model.
The planet signal of a single pixel over time is a positive transit curve modelled by simulating the planet trajectory due to the sky rotation in pupil-tracking observations. The systematic noise, including the stellar diffraction pattern, speckle, instrument scattered light, wind-driven halo, sky background, and other atmospheric effects, usually impacts a large region on the detector. TRAP reconstructs the systematic noise of the assumed planet transiting region, named the reduction region, from reference pixels that share the same underlying sources of systematic noise as the reduction region.

Because gvAPP180 produces two coronagraphic PSFs simultaneously, TRAP can select non-local pixels not only from the same PSF, which is the case for the conventional coronagraphic PSF, but also from the other PSF, which doubles the total number of available reference pixels. Thus TRAP can make the full utilisation of and explore different pixel combinations of the gvAPP180 coronagraphic data, instead of discarding partial information as several previously implemented spatial algorithms did, such as cADI. The previous algorithms based on the spatial symmetry of the two PSFs by \cite{Otten2017} and \cite{Sutlieff2021} are limited by the asymmetric noise in the two PSFs. 
TRAP may break this limitation if the noise is temporally correlated.
Besides, applying a temporal algorithm that uses the causal information between pixels, such as TRAP, can reveal the temporal correlation of the systematic noise in high-contrast imaging, which in return can benefit the improvement of adaptive optics (AO) systems and coronagraphs such as vAPPs.


In this work, we apply the TRAP algorithm on two types of vAPP coronagraphic PSFs produced by dgvAPP360 and gvAPP180. Our main focus is to investigate the selection of reference pixels for vAPP coronagraphic data, especially for gvAPP180. We compare the photometric results obtained by TRAP with those obtained by previously implemented algorithms that demonstrated the best results on these datasets.

In Sect.~\ref{sec:vAPP datasets} we provide a detailed description of the data employed in this study. We explain our methods of implementing TRAP on the dgvAPP360 and gvAPP180 coronagraphic PSFs, including the selection of reference pixels, in Sect.~\ref{sec:algorithms}. We also reproduce the spatial algorithm with the best planetary detection results on the same dataset
In Sect.~\ref{sec:Results} we present the results of different methods and compare their photometric measurements of the planets. We discuss the results in Sect.~\ref{sec:Discussion} and summarise this work in Sect.~\ref{sec:Conclusions}.

\section{vAPP datasets}
\label{sec:vAPP datasets}
We used three on-sky vAPP coronagraphic datasets summarised in Table~\ref{Table_datasets}. They are the first few science observations obtained by the vAPP coronagraphs. Two of them contain real companions that can be used to compare the performance of different algorithms.
\subsection{dgvAPP360 dataset of HR~8799}
The first one is the IFS dataset of HR~8799 planetary system obtained by \cite{Doelman2022} from LBT with the $L/M-$band (3--5\,$\mu$m) InfraRed Camera (LMIRcam) \citep{Skrutskie2010} on 18 September 2019. One dgvAPP360 was installed on LMIRcam and is also compatible with the Arizona Lenslets for Exoplanet Spectroscopy (ALES), an adaptive optics integral field spectrograph working at 2--5\,$\mu$m \citep{Skemer2015, Stone2022}. The field of view (FoV) of ALES for this dataset is 2\farcs5 $\times$ 2\farcs5 and covers the three innermost planets of the HR~8799 system (from furthest to closest): c, d, and e. The plate scale of one spaxel is 35\,mas. The planetary system was observed in the $L$-band from 2.74 to 4.29\,$\mu$m with a spectral resolution of $\sim$35, resulting in 100 wavelength channels. There are 1300 science frames in each channel with an integration time of 3934 milliseconds for each frame. There are also 99 background frames with the same integration time per frame as the science frames: the first 13 frames were observed after the first 100 science frames and all the last 86 frames were taken after the last science frame. The total observation time is 1.7\,hr with a field rotation angle of 85\fdg63. The seeing was 0\farcs8 -- 1\farcs1 during the observation.

\begin{figure}[htbp]
    \centering 
    \includegraphics[width=0.4\textwidth]{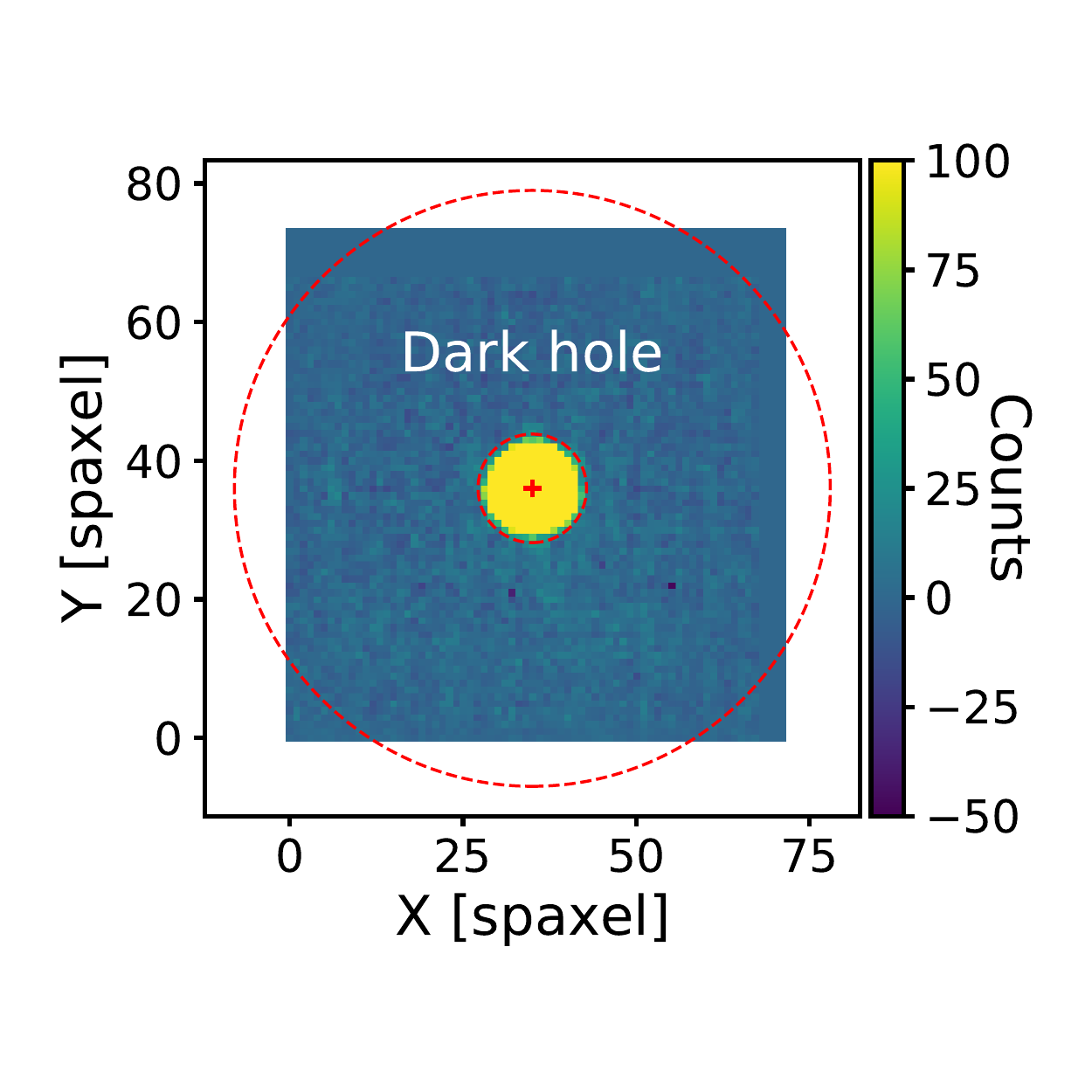}
    \caption{On-sky coronagraphic PSF from the dgvAPP360 mounted on LBT/ALES at 3.73\,$\mu$m. The IWA and OWA are shown by the two dashed circles in red. The area between them is the dark hole of the PSF where the planet is expected to be located. The image size of ALES is smaller than the OWA. The colourbar shows the flux intensity after background subtraction.}
    \label{fig:dgvAPP360_PSF}
\end{figure}

The coronagraph used in the HR~8799 dataset is the dgvAPP360 that has two liquid-crystal layers with specifically designed phase patterns that can suppress the stellar flux and form a circular dark zone surrounding the core of the central star \citep{Doelman2020}. The dark zone covers a $360^{\circ}$ angle around the central star and the resulting PSF is similar to PSFs of the focal plane coronagraphs. Because of the double-grating design, the dgvAPP360 does not split the star light to the right and left circular polarisation states as the previously designed gvAPP180 does and produces only one PSF.
The working angle of this dgvAPP360 is from 2.7\,$\lambda/D$ to 15\,$\lambda/D$ and is aimed to minimise the stellar diffraction halo close to the stellar PSF core. The inner working angle (IWA) and the outer working angle (OWA) of vAPPs are defined as the smallest separation and the largest separation where the throughput of the planet reaches half of the maximum throughput of the region where the contrast reaches the target contrast \citep{Doelman2021}. The raw contrast can reach $10^{-4}$ in the dark zone in $L$-band. A dgvAPP360 coronagraphic PSF example of the HR~8799 dataset is shown in Fig.~ \ref{fig:dgvAPP360_PSF}.

\subsection{gvAPP180 datasets of HR~2562 and Altair}
Two datasets come from the 6.5\,m Magellan Clay telescope with the adaptive optics (MagAO) system at Las Campanas Observatory, Chile \citep{Close2010, Morzinski2014}. They were collected by the same instrument at 3.94\,$\mu$m with a band width of\,90 nm: the Clio2 Narrow near-IR camera \citep{Sivanandam2006, Morzinski2015} and the gvAPP180 coronagraph \citep{Otten2017}. The detector has a size of 1024 $\times$ 512 pixels with a pixel scale of 15.85\,mas/pixel. One dataset was obtained on 07 February 2017 \citep{Sutlieff2021} targeting at the F5V star HR~2562 which hosts a brown dwarf companion, HR~2562~B, at a separation of 643.8 $\pm$ 3.2\,mas \citep{Maire2018}. This dataset used an ABBA nodding pattern: the stellar PSF was alternately imaged on the left (A) and right (B) half of the detector. There are 354 cubes with 10 frames in each cube, 3540 frames in total (A: 1770 and B: 1770), and each frame has an integration time of 4\,s with a total on-target integration time of 3.93\,hr. The derotator was on for the first 45$\%$ of the frames and switched off for the remaining part of the observation, producing a total field rotation angle of 42\fdg29. The seeing was 0\farcs5 -- 1\farcs3 during the observation. The target of the other dataset is a much brighter A-type star, \object{Altair}, one of the first on-sky datasets obtained using this gvAPP180. The science PSFs were imaged only on the left half of the detector without any nodding pattern. 300 science data cubes were taken, each with 20 subframes. The integration time of each subframe is 1\,s and the total on-target observation time is 6000\,s. The telescope derotator was off during observation, producing a total field rotation angle of 42\fdg05. There are also 20 sky frames with the same integration time as the science frame, 1\,s per frame. The star PSFs of the science frames were saturated and another 7 science data cubes (7 $\times$ 20 = 140 subframes) and 4 sky frames were taken as the calibration data. The integration time per calibration frame was 280\,milliseconds. The seeing was 0\farcs4 -- 0\farcs7 during the observation.

\begin{figure}[htbp]
    \centering 
    \includegraphics[width=0.5\textwidth]{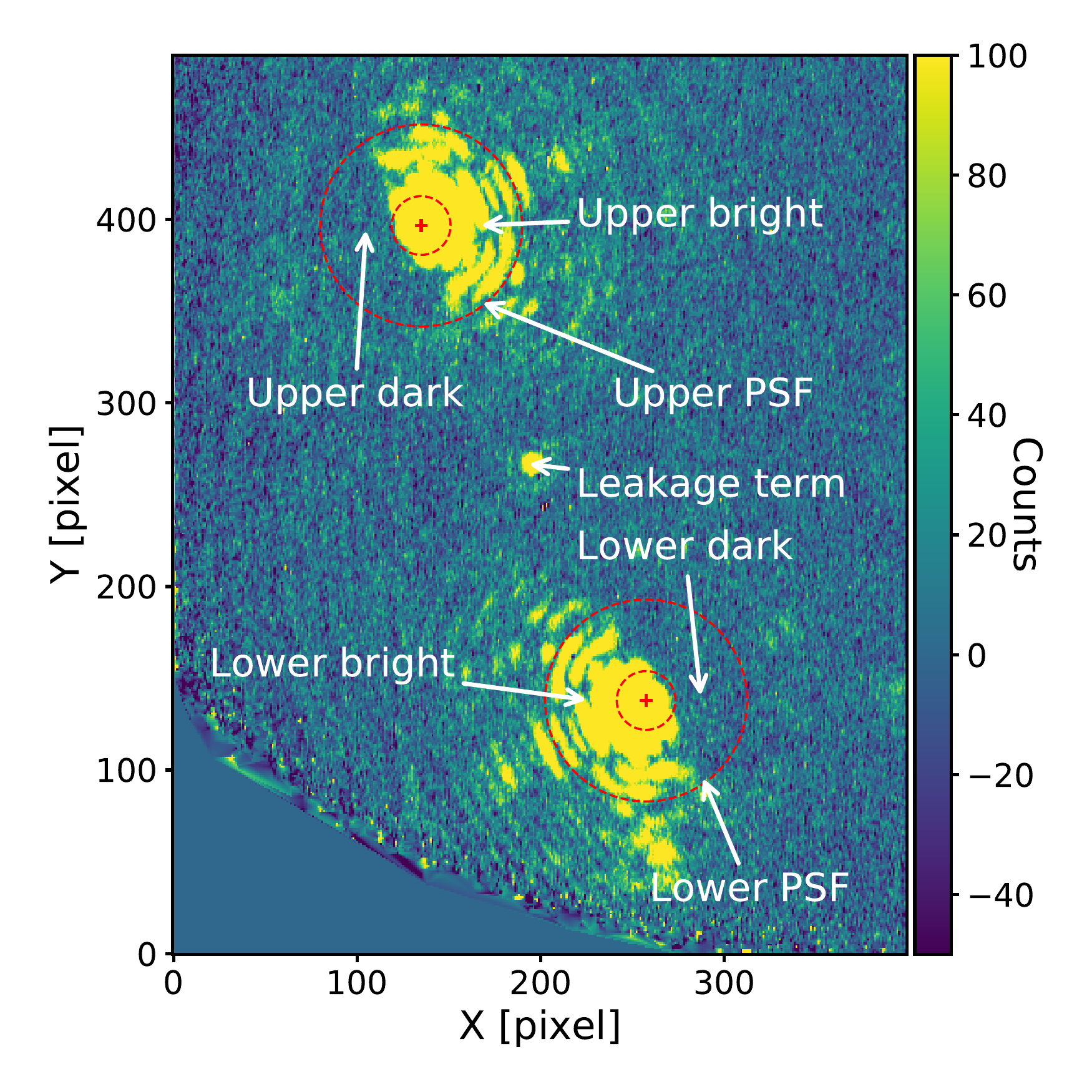}
    \caption{On-sky coronagraphic PSF from the gvAPP180 mounted on MagAO/Clio2 at 3.94\,$\mu$m. One PSF (upper PSF) is above another PSF (lower PSF) with a leakage term in the middle. The red circles in dashed lines show the IWA and OWA. The dark holes of the two PSFs complement the FoV of each other. The four different regions of the two PSFs we defined in this work are marked with white text. The colourbar shows the flux intensity after background subtraction.}
    \label{fig:gvAPP180_PSF}
\end{figure}

This gvAPP180 coronagraph produces two coronagraphic PSFs by splitting the two circular polarisation modes of the star light \citep{Otten2017}. There is also a leakage term between the two PSFs, an intact stellar PSF without any dark holes. The intensity of the leakage term is faint and we did not make use of it in this work. All objects in the field of view receive the same phase adjustment by the gvAPP180 coronagraph. Hence the planetary PSF has the same shape as the stellar PSF in each coronagraphic PSF. If there is a planet located in the dark hole of one PSF, it also exists at the same location in the bright side of the other PSF.
The working angle of this vAPP is between 2 and 7\,$\lambda/D$ and works from 2 to 5\,$\mu$m. 
A gvAPP180 coronagraphic PSF example of MagAO/Clio2 is shown in Fig.~\ref{fig:gvAPP180_PSF}.
Each PSF has a half dark side where the starlight is significantly suppressed and a half starlight-unsuppressed bright side. Thus we have four different regions from the two opposite PSFs (upper PSF and lower PSF): upper dark, upper bright, lower dark, and lower bright.

\begin{table*}[h!]
\centering
\caption{vAPP datasets.}
\begin{tabular}{l l l l l l l l}
\hline\hline
 Dataset & vAPP & Pixel scale (mas/pixel) & Wavelength ($\mu$m) & DIT\tablefootmark{a} (s) & Drift\tablefootmark{b} & Rotation\tablefootmark{c} & Companion\tablefootmark{d}\\
\hline
HR~8799 & LBT/ALES/dvAPP360 & 35.00 & 2.74--4.29 & 3.9 & small & 85\fdg63 & cde\\
HR~2652 & MagAO/gvAPP180 & 15.85 & 3.94 & 4.0 & large & 42\fdg29 & B\\
Altair &  MagAO/gvAPP180 & 15.85 & 3.94 & 1.0 & large & 42\fdg05 & -\\
\hline
\end{tabular}
\tablefoot{
\tablefoottext{a}{Detector integration time of each frame.}
\tablefoottext{b}{Stellar PSF drift in frames.}
\tablefoottext{c}{Field rotation.}
\tablefoottext{d}{Known companions in the field of view.}
}
\label{Table_datasets}
\end{table*}

\subsection{Object property}
HR~8799 hosts four giant, wide-orbit planet-mass objects discovered by direct imaging \citep{Marois2008,Marois2010}. The FoV of our dataset only covers the three inner planets, HR~8799~cde. HR~8799 d and e have a spectral type of L6-L8, and HR~8799 c is likely to have a later spectral type \citep{Zurlo2016,Bonnefoy2016}. HR~2562~B is a substellar companion orbiting HR~2562 and has a spectral type of L7$\pm 3$ \citep{Konopacky2016}. Altair is a bright A type star without any known companions. Table~\ref{Table_objects} lists the properties of the central stars and their companions from the literature. 

\begin{table*}[h]
\centering
\caption{Properties of the observed objects from the literature.}
\begin{tabular}{l l l l}
\hline\hline
Stellar & Mass ($M_{\odot}$) & Spectral Type  & \\
\hline
HR 8799 & 1.47 $\pm$ 0.30 [1] & F0+VkA5mA5 [2] &  \\
HR 2562 & 1.37 $\pm$ 0.02 [3] & F5V [4] & \\
Altair & 1.79 $\pm$ 0.02 [5] & A7V [2] &\\
\hline
Companion & Mass ($M_{\rm J}$) & Contrast ($\times 10^{-4}$) & Separation $\rho$ (mas), $\phi$ (deg) \\
\hline
HR 8799 c & 7.8 $\pm$ 0.5 [6] & $1.67^{+0.13}_{-0.12}$ [7] & 947 $\pm$ 7, 327.83 $\pm$ 0.44 [8]\\
HR 8799 d & 9.1 $\pm$ 0.2 [6] & $1.92^{+0.17}_{-0.15}$ [7] & 654 $\pm$ 7, 217.66 $\pm$ 0.81[8]\\
HR 8799 e & 9.6 $^{+1.9}_{-1.8}$ [11] & $1.96^{+0.29}_{-0.25}$ [7] & 381 $\pm$ 7, 272.48 $\pm$ 0.72 [8]\\
HR 2562 B & 32 $\pm$ 14 [3] & 3.05 $\pm$ 1.00 [9] & 635 $\pm$ 3, 297.51 $\pm$ 0.28 [10] \\
\hline
\end{tabular}
\tablebib{: [1]: \cite{Gray1999}; [2]: \cite{Gray2003}; [3]: \cite{Mesa2018}; [4]: \cite{Gray2006}; [5]: \cite{Monnier2007}; [6]: \cite{dziewski2020}; [7]: Keck/NIRC2-2012, 3.8\,$\mu$m, \cite{Currie2014}; [8]: 2014.Dec, \cite{Apai2016}; [9]: 3.94\,$\mu$m, \cite{Sutlieff2021}; [10]: 2017.Feb, \cite{Maire2018}. [11]: \cite{Brandt2021}.}
\label{Table_objects}
\end{table*}

\section{Post-processing algorithms}
\label{sec:algorithms}
The main goal of this work is to implement and test TRAP on vAPP coronagraphic data. The unconventional coronagraphic PSFs of vAPP have more flexibility in choosing non-local training data for the systematics model compared to conventional ones. This approach provides a unique opportunity to investigate the noise spatial and temporal correlations in vAPP coronagraphic PSFs and optimise the planet detection using a combination of training data. Additionally, this study may shed light on the underlying systematic noise and contribute to the development of future unconventional PSF designs.
We extend the TRAP pipeline from \cite{Samland2021} to support the dgvAPP360 and gvAPP180 coronagraphic data and increase options in reference pixel selection. We also reproduce the previously implemented algorithms on vAPP coronagraphic data. Algorithms implemented on all datasets are described below. 

\subsection{Algorithms implemented on the dgvAPP360/HR~8799 dataset}
\label{sec: algorithms on dgvAPP360}
The coronagraphic PSF of dgvAPP360 is a single symmetric PSF with a circular dark zone. We can directly apply ADI-based algorithms to this dataset as a regular pupil-tracking dataset without additional processing of the PSFs. Two algorithms are implemented on this IFS dataset and each wavelength channel is reduced independently.

Custom speckle subtraction algorithm (Custom): The sparsely sampled background frames, PSF drift ($\sim$ 2\,spaxels or 70\,mas), and global patterns probably introduced by the ALES pipeline make it very challenging to apply standard algorithms such as ADI or PCA to this dataset.
Therefore, \cite{Doelman2022} design a custom spatial-domain method based on ADI and PCA to reduce the data. We follow their description to implement it on this dataset. 
Every four science frames were binned. Ten principal components of the 99 background frames were fitted to and subtracted from the science frames. The residuals were combined as a stellar PSF model. 
Then the stellar PSF model was scaled and subtracted from the original science frames. Afterwards, the star-subtracted frames were divided into eight groups in chronological order. For frames in each group, the background was removed by fitting ten principal components of frames that are at least 15-minute away from this group ($\sim$0.85\,$\lambda/D$ for HR~8799~e) to the frame and subtracted from it.
Then a low-order polynomial fitting along every row and column was used to remove the global structure, which varies between frames. The residual frames were aligned to north and then median combined. To remove the residual low-frequency noise, the final image was convolved with a Gaussian 2D kernel with a standard deviation of five pixels and subtracted by it.

\begin{figure}[htbp]
    \centering 
    \includegraphics[width=0.5\textwidth]{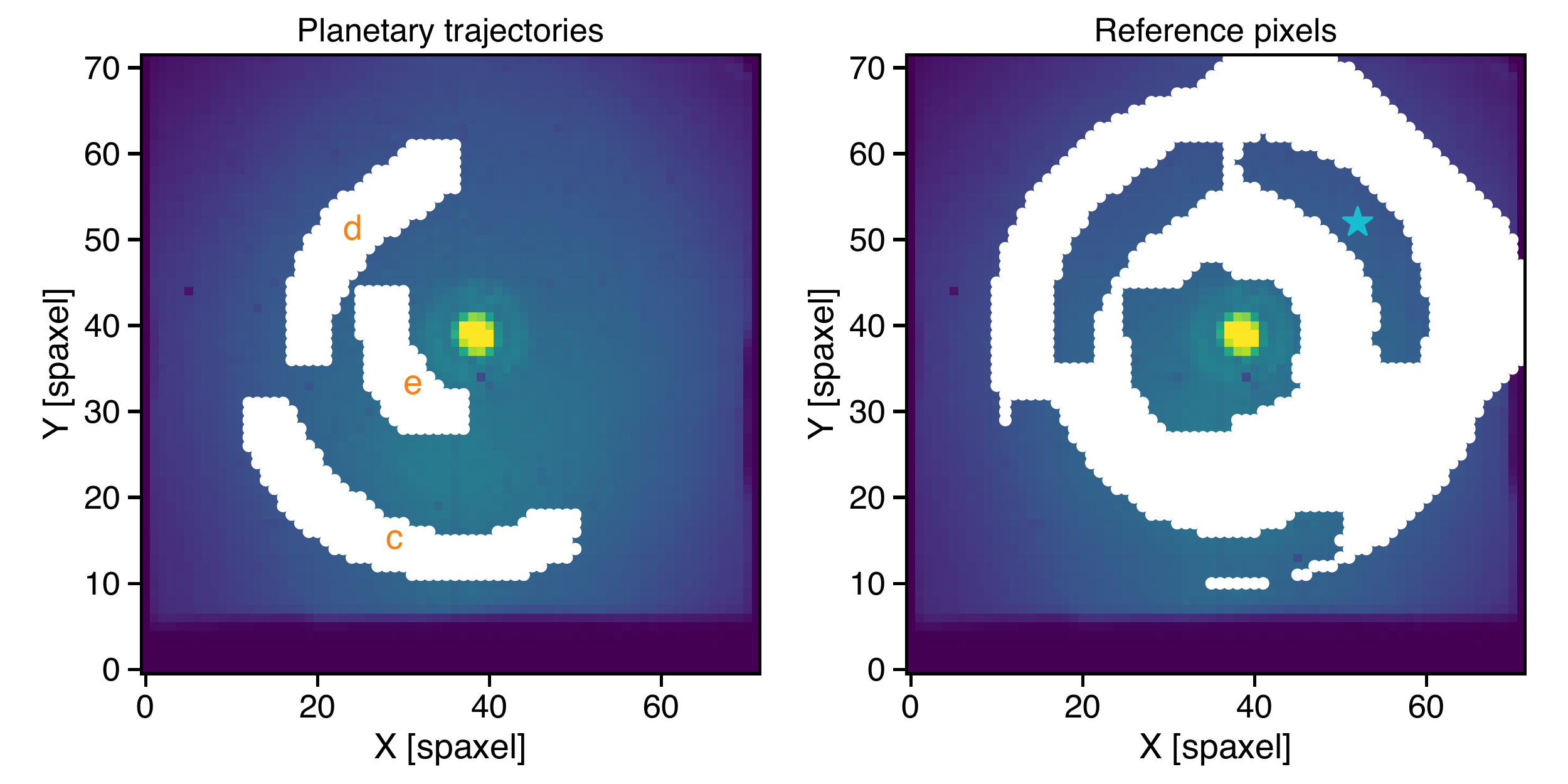}
    \caption{Planetary trajectories and reference pixel selection of the dgvAPP360 dataset. Left panel: planetary trajectories of HR~8799~cde. The image drift during the observations is also included in the trajectory. Right panel: reference pixels for a supposed position (cyan asterisk) used in TRAP, as shown by the white pixels. The known planet signals are masked.}
    \label{fig:vAPP360_reference}
\end{figure}

TRAP: TRAP is a forward model that simulates planet positions in the original dataset and reconstructs all shared systematics, including the background, stellar speckles and other noise together with the planet signal in science frames using a causal time series linear regression model. A set of basis vectors are constructed by the principal analysis of the light curves of reference pixels that are impacted by the same noise sources as the reduction region. Therefore, the problems of this dataset are solvable in TRAP. Furthermore, because TRAP is a temporal algorithm, it works better at shorter integration time, for instance, a few seconds when the speckle temporal correlations are not averaged out \citep{Samland2021}.
Therefore, the 1300 science frames were directly put into TRAP without any binning operation and pre-alignment. We used the same PSF model as the custom algorithm. For the reduction region, a PSF model with a radius of 1.1\,$\lambda/D$ was used to build the planet trajectory and the transit light curve of each pixel. 
For the reference pixel selection, we adopted a similar selection rule as in \cite{Samland2021}: an annulus at the same separation of the reduction region, pixels surrounding the reduction region, and symmetrical pixels of the reduction region on the opposite side of the star. Pixels affected by the real planet signal were excluded from reference pixels, which is the known companion mask. The annulus width of the reference pixels was set to be 20 pixels to ensure the number of reference pixels was enough to reconstruct the systematic noise. Figure~\ref{fig:vAPP360_reference} shows the actual planetary trajectories due to field rotation and an example of the reference pixel selection of a supposed planet position. We also excluded reference pixels within the IWA when reducing regions outside the IWA, which resulted in a higher S/N.

\subsection{Algorithms implemented on the gvAPP180/HR~2562 dataset}
\label{sec:algorithms on gvAPP180}
The D-shaped PSFs of gvAPP180 require specially designed data reduction techniques different from those for conventional circular coronagraphic PSFs. \cite{Sutlieff2021} apply three techniques to the gvAPP180/HR~2562 dataset and find that cADI produces the strongest detection of the companion among the three techniques. We follow their method description and find similar results. Hence, we choose cADI as the conventional speckle subtraction algorithm to compare with TRAP.

The star drifted linearly towards the lower left of the detector when the derotator was on, about 30\,pixels (475\,mas) on the y-axis and 35\,pixels (555\,mas) on the x-axis. Centring on the PSF core, we cropped the upper and lower PSFs separately with a frame size of 170 $\times$ 170 pixels (2\farcs7 $\times$ 2\farcs7), covering the whole working region of the gvAPP180. Then we did the linear, dark frame and flat field correction. We subtracted frames of one nod from the other to remove the thermal background.
Furthermore, we binned every five frames in time to increase S/N and save computational time. The final total number of frames is 708 from the two nods.

Classical ADI: The two dark holes from the lower and upper PSFs were joined together without overlap, and then regions inside the IWA and beyond the OWA were masked. As a reference PSF, the median of all frames was subtracted. Then the residual frames were derotated to north. We took the median of the derotated frames as the final residual map of cADI.

TRAP: The large drift of PSFs in the HR~2562 dataset makes the background contain not only temporal variations but also spatial variations due to the drift, for instance, the change of patterns spatially fixed in the detector. Therefore, the background could not be well reconstructed by TRAP in the aligned images. So we put the background subtracted and aligned images into TRAP. The aligned but background unsubtracted frames were used as the variance dataset for uncertainty calculation.
Additionally, we implemented TRAP on cube A and cube B separately as they belong to different nod positions and hence very different regions in the detector, sharing few common temporal variations.
By visual inspection, cube B was more heavily influenced by the detector and instrument effects, including a dark reflection ghost on the bright side of the upper PSF and a bright spike of scattered light across the lower PSF. The median of one cube was taken as the PSF model, and thus we had an upper PSF model and a lower PSF model.
Furthermore, the flux in the dark holes of the PSF models was set to zero so that pixels with negligible planet signal from this area are not included in the reduction region (simulated by the PSF model) and can be used as reference pixels instead. Because they are next to the reduction region, they share similar noise as the reduction region and can increase the accuracy of the systematics model.
We found that a larger PSF model size had a better result for cube B due to the more diffused planetary flux in nod B. The final PSF model used to simulate the reduction region has a radius of 1.1\,$\lambda/D$ for cube A, while the radius is 2.1\,$\lambda/D$ for cube B.

\begin{figure}[htbp]
    \centering 
    \includegraphics[width=0.5\textwidth]{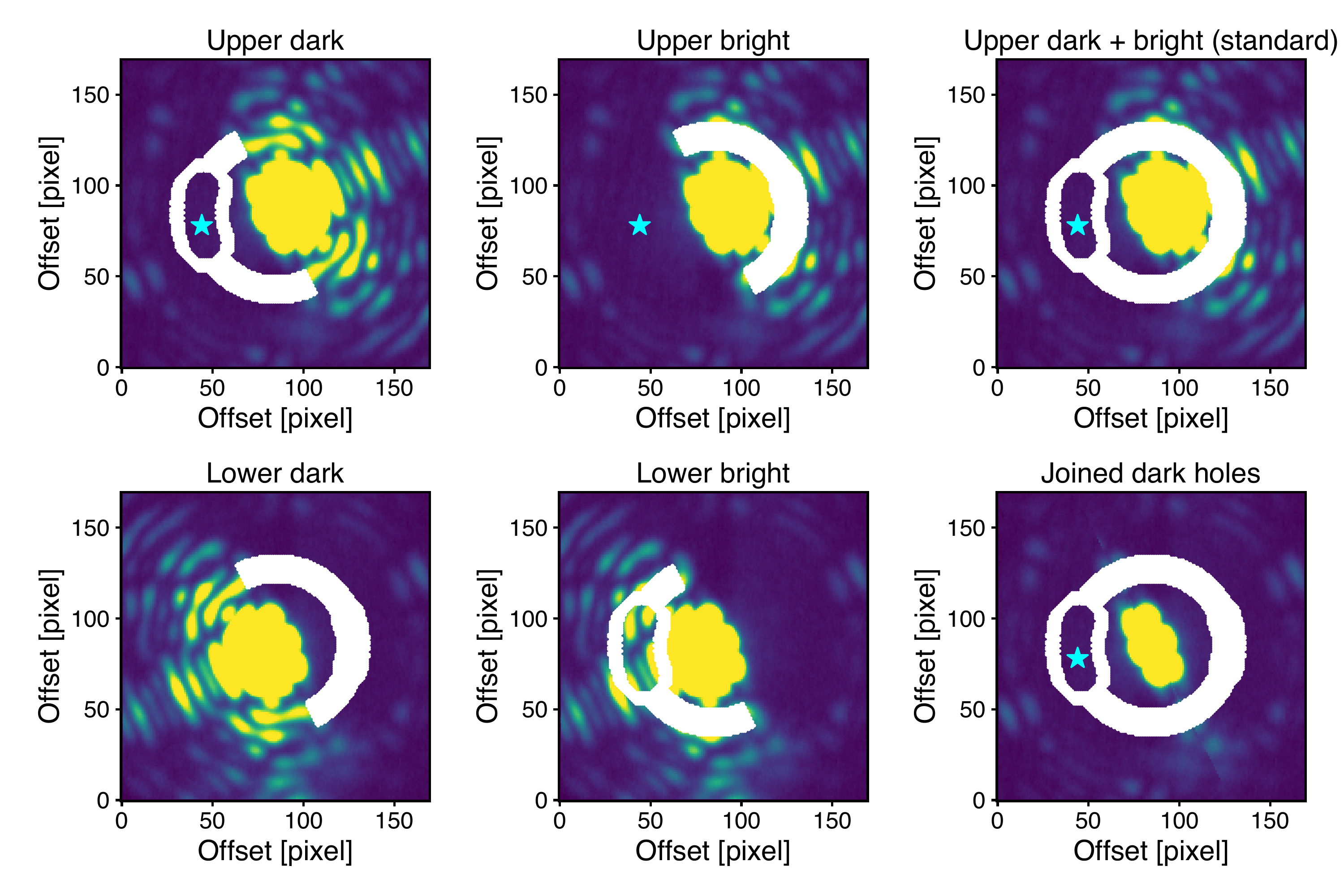}
    \caption{Six reference pixel designs for an assumed planet position (cyan asterisk) in the upper dark hole of a gvAPP180 PSF, as shown by the white pixels. Upper dark: choosing reference pixels from the dark side of the same PSF; upper bright: choosing reference pixels from the bright side of the same PSF; lower dark: choosing reference pixels from the dark side of the complementary PSF; lower bright: choosing reference pixels from the bright side of the complementary PSF; upper dark+bright: choosing reference pixels from the dark and bright sides of the same PSF; joined dark holes: choosing reference pixels from the joined dark holes.
    }
    \label{fig:gvAPP180_reference_pixels}
\end{figure}

The key question when applying TRAP for gvAPP180 is how to choose reference pixels for the reduction region. Based on the same selection rules for dgvAPP360, we added the `half annulus' option for gvAPP180: splitting the annulus into dark and bright sides. We heuristically designed six choices to select reference pixels from the four different regions of the upper and lower PSFs:
\begin{itemize}
\item Upper dark: choosing reference pixels from the dark side of the same PSF; 
\item Upper bright: choosing reference pixels from the same PSF, but only from the bright side; 
\item Lower dark: choosing reference pixels from the dark side of the complementary PSF; 
\item Lower bright: choosing reference pixels from the bright side of the complementary PSF, but the same positions as the reduction region in the upper dark hole have to be masked since they also contain planet signals; 
\item Upper dark+bright (standard): choosing reference pixels from both dark and bright sides of the same PSF, the same as for a symmetric PSF (e.g. the dgvAPP360 PSF); 
\item Joined dark holes: choosing reference pixels from the two joined dark holes.
\end{itemize}
White pixels in Fig.~\ref{fig:gvAPP180_reference_pixels} demonstrate the six reference pixel options when searching in the upper dark hole, where the companion is located. 
Pixels influenced by known companion signals in both upper and lower PSFs were excluded from the reference pixel selection. The selected pixel for demonstration in Fig.~\ref{fig:gvAPP180_reference_pixels} is also the expected companion location and thus the reduction region overlaps the known companion mask.
When reducing the position within the IWA, we used all pixels outside the reduced region within the IWA as reference pixels to ensure that the number of reference pixels is greater than the number of frames.

\section{Results}
\label{sec:Results}
\subsection{Photometric measurement and spectra comparison of the dgvAPP360/HR~8799 dataset}
\label{sec:dgvAPP360_result}
\begin{figure*}
\centering 
\resizebox{\hsize}{!}
          {\includegraphics[width=\textwidth]{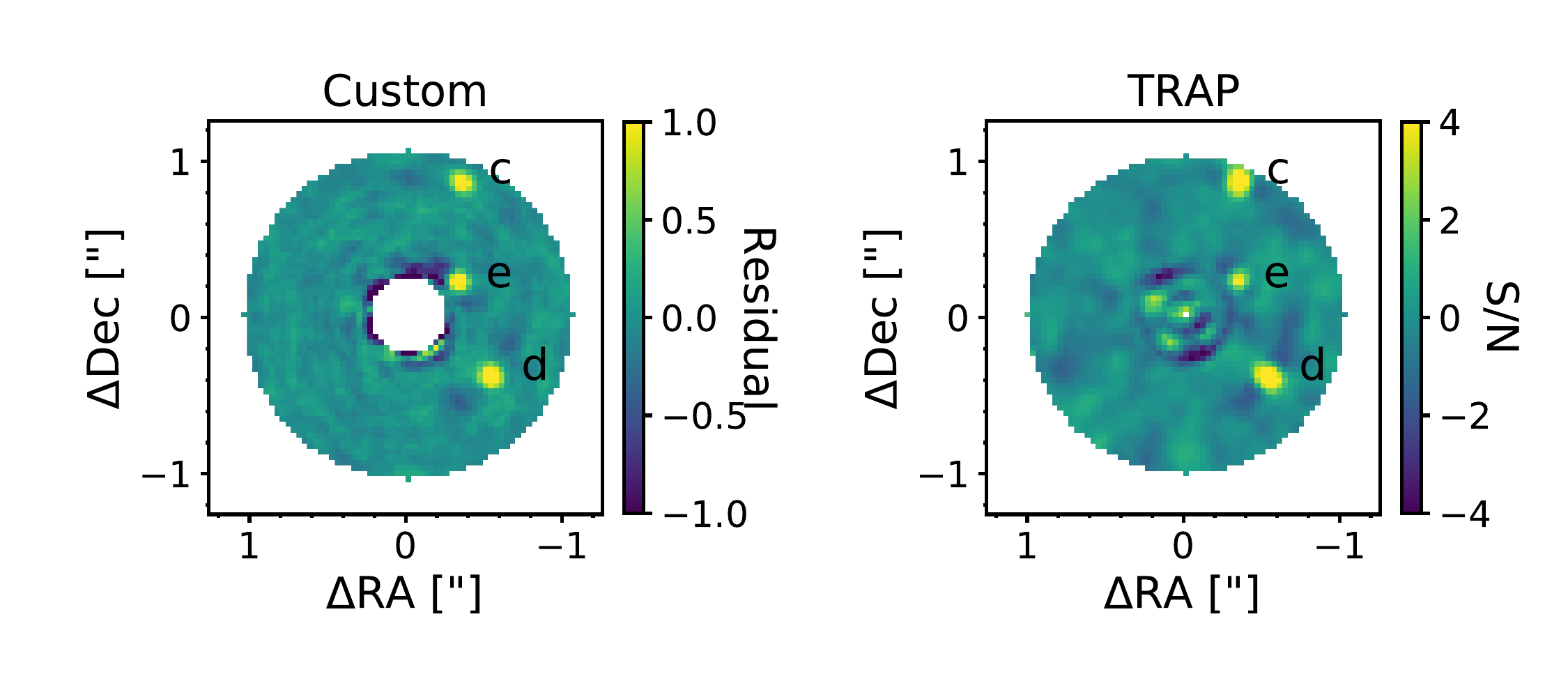}}
          \caption{HR~8799 result maps of the custom algorithm and TRAP averaged between 3.52 and 4.10\,$\mu$m. Left panel: the final residual map of the custom algorithm where the colourbar represents the residual level. Right panel: the normalised S/N map of TRAP where the colourbar represents the S/N level. The two maps should not be compared directly.
          }
 \label{Fig:dgvAPP360_CustomTrap}
\end{figure*}

The residual map of the custom algorithm and normalised S/N map of TRAP averaged in 3.52--4.10\,$\mu$m are presented in Fig.~\ref{Fig:dgvAPP360_CustomTrap}. HR~8799~cde are clearly detected by the two algorithms.
For the custom algorithm, negative planet signals were injected to extract the planet location and contrast to the star \citep{Marois2010,Lagrange2010}. The negative planet signals were injected at the assumed planet locations in each frame. Then the injected dataset was reduced by the custom algorithm. The injected planet location and contrast were optimised iteratively with the downhill simplex method \citep{Nelder1965} until the residual planet signals were minimised in the residual map \citep[e.g.][]{Stolker2019}. 
For TRAP, we measured the photometry by fitting a 2D Gaussian model to the unnormalised S/N map, which is the contrast map divided by the uncertainty map, because fitting in the normalised S/N map may slightly bias the planet location. Instead of interpolating the contrast at this position (sub-pixel precision), we then reduced the best-fit position using TRAP to obtain the contrast and uncertainty. The uncertainty was then normalised by interpolating the normalisation profile calculated from the unnormalised S/N map.

\begin{figure*}
\centering 
\resizebox{\hsize}{!}
          {\includegraphics{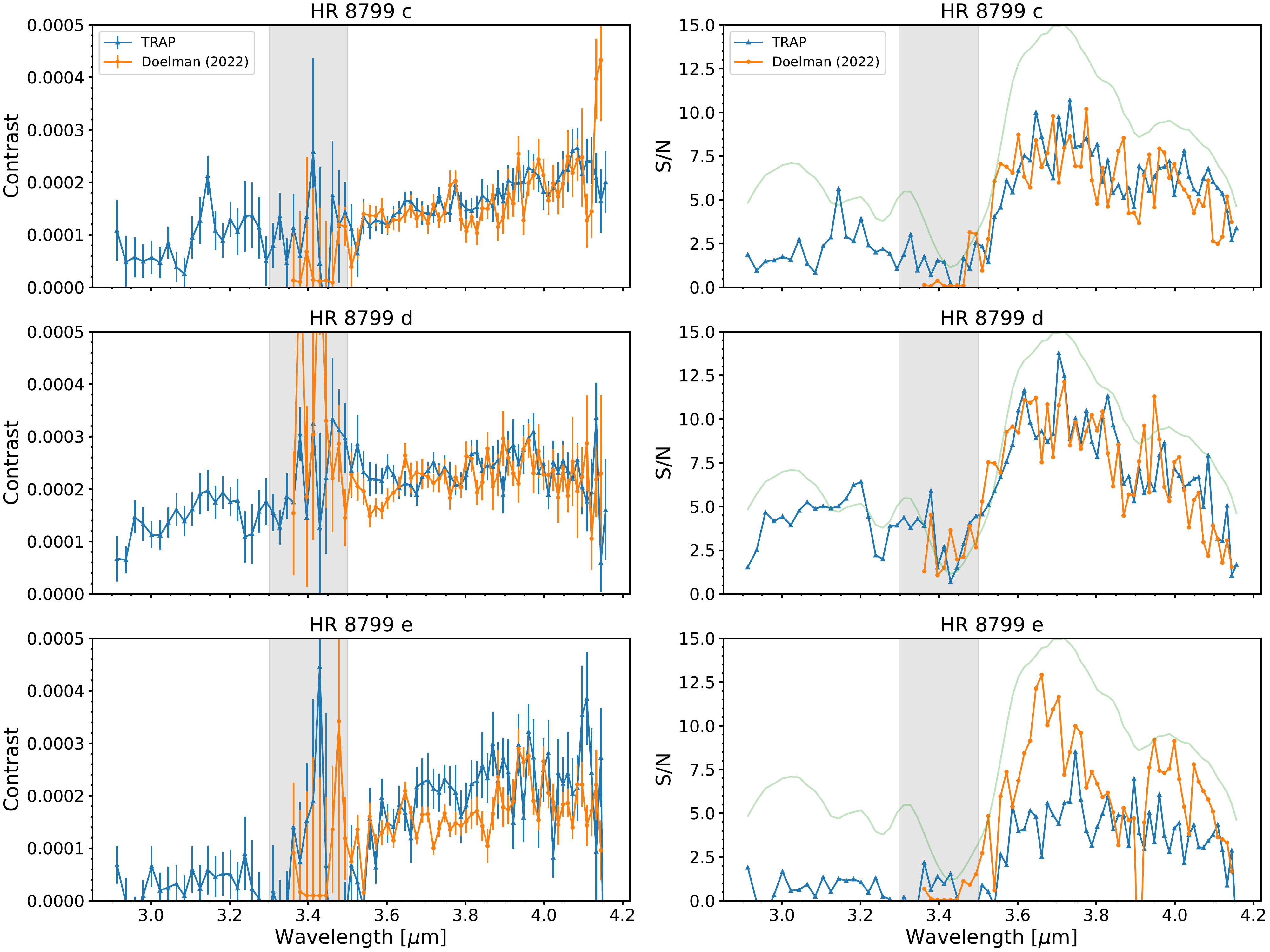}}
          \caption{Contrast spectra and S/N comparison of the custom algorithm and TRAP. The results of the custom algorithm are adopted from the published results in \cite{Doelman2022}, covering 3.36 to 4.14\,$\mu$m. Left panel: the extracted spectra of HR~8799~c, d and e by the two algorithms between 2.91 and 4.16\,$\mu$m. Right panel: the contrast S/N. The light green profile in the right panel is the star flux normalised by its maximum value between 2.91 and 4.16\,$\mu$m, which is an indicator of the transmission of the dgvAPP360 coronagraph. The grey shaded area shows the absorption band of the coronagraph, where the extracted planet signals are the worst. On average, TRAP achieves slightly higher S/N and smoother contrast spectra for HR~8799~c and d. The custom algorithm achieves higher S/N for HR~8799~e but with a larger deviation between wavelength channels than TRAP.
          }
 \label{Fig:dgvAPP360_spectraSNR}
\end{figure*}

\cite{Doelman2022} use bootstrapping to estimate error bars for the retrieved contrasts. Their contrast spectra have higher S/N and smaller scatter than our version of the custom algorithm due to the difference in detailed implementation.
We adopt their results on behalf of the custom algorithm to compare with the results of TRAP.
The stellar PSFs in the blue and red edges of the IFS wavelength channels are blurred, which might be caused by cross-talk between wavelength channels. We discarded these channels and reduced the remaining 81 channels covering 2.91 to 4.16\,$\mu$m with TRAP. \cite{Doelman2022} only calculate the uncertainty for each channel from 3.36 to 4.14\,$\mu$m where the dgvAPP360 has high transmission. For wavelength shorter than 3.36\,$\mu$m, they combine multiple channels to obtain two representative points. Therefore, we compare the results only in the common spectral range.
The planet-star contrast spectra extracted with the two methods are presented in Fig.~\ref{Fig:dgvAPP360_spectraSNR}. They have similar profiles and agree with each other within one sigma in most wavelength channels.  
The considerable scatter and low S/N between 3.3 and 3.5\,$\mu$m are due to the absorption band of the dgvAPP360 coronagraph \citep{Otten2017,Doelman2022}. 
The contrasts (here refers to planetary brightness relative to the star) of the three planets increase towards longer wavelengths since the planets are brighter at longer wavelengths. 
The S/N curves of the three planets are consistent with the dgvAPP360 coronagraph transmission profile, peaking at 3.65--3.75\,$\mu$m.
In Table~\ref{Table_HR8977} we present the S/N and contrast averaged between 3.52 and 4.10\,$\mu$m where the transmission of the dgvAPP360 is the highest.
In this range, TRAP achieves slightly higher S/N and less noisy contrast spectra for HR~8799~c and d than the custom algorithm.
For both methods, HR~8799~d is brighter than HR~8799~c and has a higher S/N, which is consistent with the literature contrast in the $L^{'}$ band in Table~\ref{Table_objects}.
However, compared to the outer two planets, the contrast spectrum of HR~8799~e has stronger high-frequency noise by both methods. Though the custom algorithm achieves a higher S/N for HR~8799~e compared to TRAP on average, the deviation of its S/N between wavelength channels is also larger. This indicates some inconsistent noise at small angular separations between different wavelength channels.

\begin{table}[h!]
\centering
\setlength{\belowcaptionskip}{10pt}
\caption{The averaged S/N and planet-star contrast of HR~8799~c, d and e extracted by the custom algorithm from \cite{Doelman2022} and TRAP between 3.52 and 4.10\,$\mu$m. The uncertainty is calculated as the standard deviation between wavelength channels.}
\begin{tabular}{l l l l}
\hline\hline
 & HR~8799~c & HR~8799~d & HR~8799~e \\ 
\hline
S/N & & & \\
TRAP & 6.67 $\pm$ 1.63 & 8.02 $\pm$ 2.12 & 4.13 $\pm$ 1.57\\
Doelman2022 & 6.36 $\pm$ 1.70 & 7.84 $\pm$ 2.42 & 6.78 $\pm$ 3.15\\
\hline
Contrast ($\times 10^{-4}$) & & & \\
TRAP & 1.71 $\pm$ 0.40 & 2.34 $\pm$ 0.28 & 2.03 $\pm$ 0.72\\
Doelman2022 & 1.63 $\pm$ 0.44 & 2.22 $\pm$ 0.34 & 1.69 $\pm$ 0.50\\
\hline
\end{tabular}
\label{Table_HR8977}
\end{table}

\subsection{The gvAPP180/HR~2562 dataset}
We reduced the two MagAO/gvAPP datasets of HR~2562 and Altair using cADI and TRAP. Because HR~2562 has a known companion, we mainly used this dataset to explore the algorithms for the gvAPP180 datasets. The Altair dataset is used for further analysis of noise temporal correlations in Sect.~\ref{sec:Discussion}.

\subsubsection{Reference pixel selection in TRAP for gvAPP180}
\label{sec:referenceTRAPgvAPP}
At first, we introduce two definitions in the TRAP detection map resulting from the dark and bright sides of the gvAPP180 PSF: dark region and transition region.
When considering the planet trajectory in the pupil tracking dataset of gvAPP180 projected into a single PSF, as in Fig.~\ref{fig:gvAPP180_PSF}: it can either fall completely into the dark hole or cross both the dark and bright sides.
If it falls completely on the bright side, this is equivalent to falling completely into the dark hole of the opposite PSF.
We define the `dark region' in the detection map as the area where all simulated planet trajectories fall only into the dark hole. As such there are upper and lower dark regions in the detection map that correspond to the dark holes of the upper and lower PSFs, respectively. The dark region is always smaller than the dark hole and should be distinguished from the `dark hole'. The rest of the detection map is referred to as the `transition region', where all simulated planet trajectories fall into both the dark and bright sides of the PSF. The transition region calculated from the upper and lower PSFs covers the same area in the detection map. It spans twice the parallactic angle range. For the transition region, we selected reference pixels from the same PSF using the standard choice as used for the dgvAPP360 PSF in Sect.~\ref{sec: algorithms on dgvAPP360}.

\begin{figure*}
\centering 
\resizebox{\hsize}{!}
          {\includegraphics{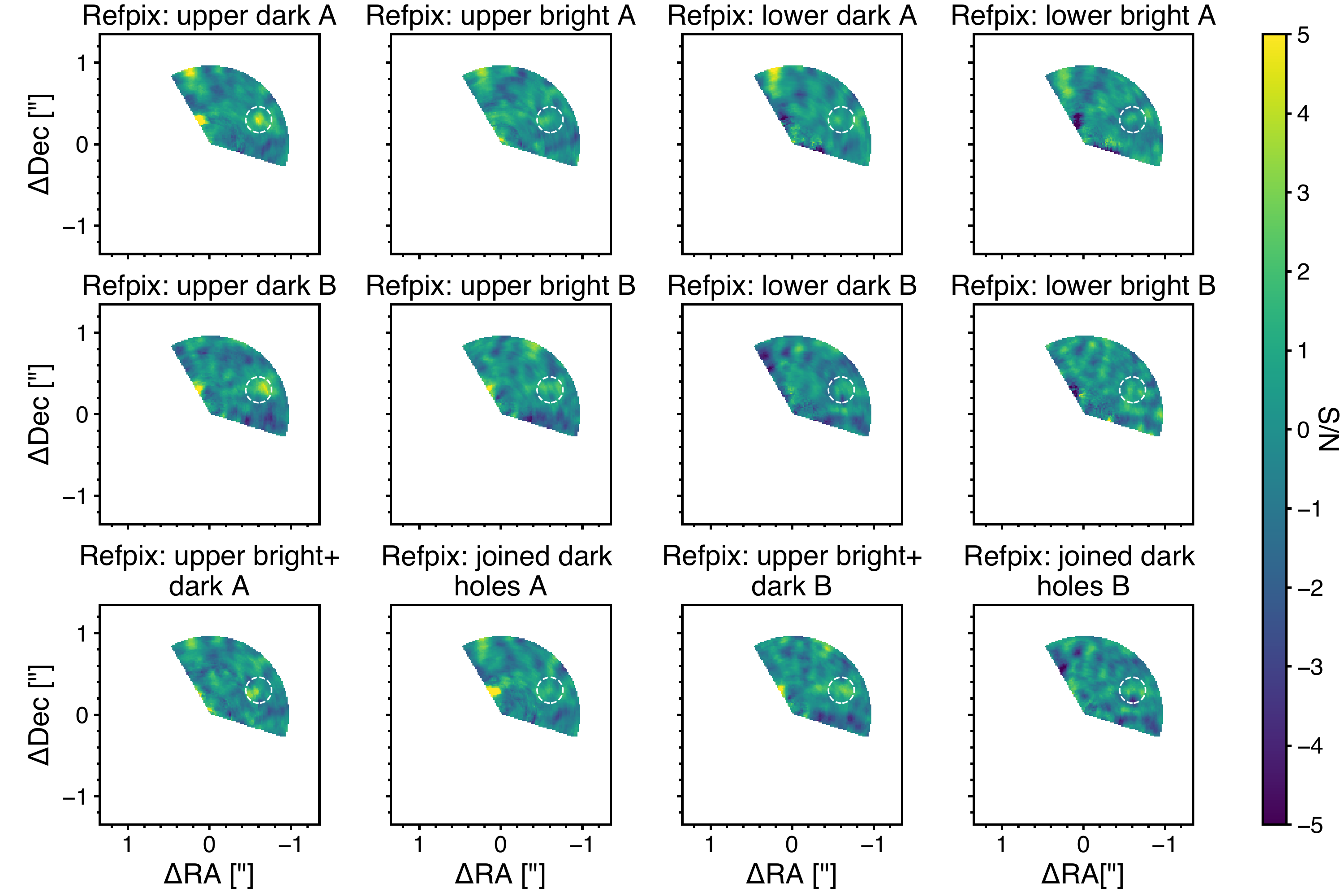}}
          \caption{S/N detection maps of the upper dark region of the gvAPP180/HR~2562 dataset reduced with the six reference pixel choices in TRAP. Cube A and B were reduced separately. The location of HR~2562~B is highlighted by the white circles. North is up and east is left. The companion is clearly detected in both cube A and B when selecting reference pixels only from the dark hole in the upper PSF, where the companion is located.
          }
 \label{Fig:gvAPP180_ABSNRmap}
\end{figure*}

The location of HR~2562~B is in the middle of the upper dark region in the detection map. We reduced the upper dark region with the six reference pixel selections demonstrated in Fig.~\ref{fig:gvAPP180_reference_pixels}. The reference pixel annulus had a width of 15 pixels and the known companion HR~2652~B was masked by a mask with a radius of 11\,pixels ($\sim$ 1.5\,$\lambda/D$). The 12 S/N maps of the upper dark regions of cube A and B are presented in Fig.~\ref{Fig:gvAPP180_ABSNRmap}. The S/N maps were normalised by a three-pixel-wide annulus in the upper dark region as a function of separation. We found that selecting reference pixels solely from the upper dark hole allowed for a clear detection of the companion in cube A and B. The detection was more pronounced in cube A than in cube B due to the latter's worse image quality.
The bright patches visible on the left edge of the upper dark region in the detection map of cube A and B are due to the starlight contamination from the bright side of the PSF. These features also appear in the residual map of cADI both in our reduction and that of \cite{Sutlieff2021} (see their Fig.~2). But they do not invalidate the detection of HR~2562~B at the expected location based on \cite{Maire2018}.

The above result suggests that we should only choose reference pixels from the same dark hole when reducing the dark region of HR~2652. Hence, we reduced the upper dark region with reference pixels from the upper dark hole and the lower dark region with reference pixels from the lower dark hole. To generate a synthetic detection map, we used the weighted average \citep{Samland2021} to combine the contrast $\omega$ and its uncertainty $\sigma$ from the two independent reductions of the transition region. For each pixel in the transition region:
\begin{equation}
\omega=\frac{\omega_{1} \sigma_{\omega_1}^{-2} + \omega_{2} \sigma_{\omega_2}^{-2}} {\sigma_{\omega_1}^{-2} + \sigma_{\omega_2}^{-2}}, \quad \sigma=\sqrt{(\sigma_{\omega_1}^{-2} + \sigma_{\omega_2}^{-2})^{-1}},
\label{eq:weighted}
\end{equation}
where $\sigma_1$ and $\sigma_2$ are the preliminary uncertainty before normalisation. The normalisation for the synthetic S/N map ($\omega / \sigma$) was calculated by dividing the dark region by its robust standard deviation within the dark part of the annulus and dividing the transition region by its robust standard deviation within the transition part of the annulus as a function of separation.

\subsubsection{TRAP and cADI comparison for gvAPP180}
The final S/N map of TRAP for the HR~2562 dataset is the weighted combination of the two normalised synthetic S/N maps from cube A and cube B also with Eq.~\ref{eq:weighted}. 
To measure the astrometry of HR~2562~B in the TRAP reduction, we fitted a 2D Gaussian model to the weighted combined contrast map. The photometric measurement of the companion was obtained by reducing the optimised position from cube A and cube B using TRAP, and then by combining the contrasts and uncertainties weightedly.
For cADI we used aperture photometry and cross-correlation to calculate the S/N map \citep{Ruffio2018}.
To create a flux map, we cross-correlated the residual map of cADI with an aperture having a diameter of seven pixels ($\sim$FWHM). Each pixel in the flux map contains the total flux within that aperture in the residual map.
The S/N map was obtained by normalising the flux map by its standard deviation within a seven-pixel-wide annulus as a function of separation.
The transition region of cADI should also span twice the parallactic angle range, the same as the transition region of TRAP. However, there are still lots of residual noise patterns at the border between the transition region and dark region due to the speckle near the joined boundary of the upper and lower dark holes. Therefore we enlarged the transition region to include them. The transition region and dark region in cADI's S/N map were also normalised separately by the part of annulus in their own area as a function of separation, the same as in TRAP. The companion contrast and position of cADI were also extracted by the negative planet injection method in the residual map. The S/N and error of the contrast were estimated by interpolating the S/N map at the optimised planet position.

\begin{figure*}
\centering 
\resizebox{\hsize}{!}
          {\includegraphics{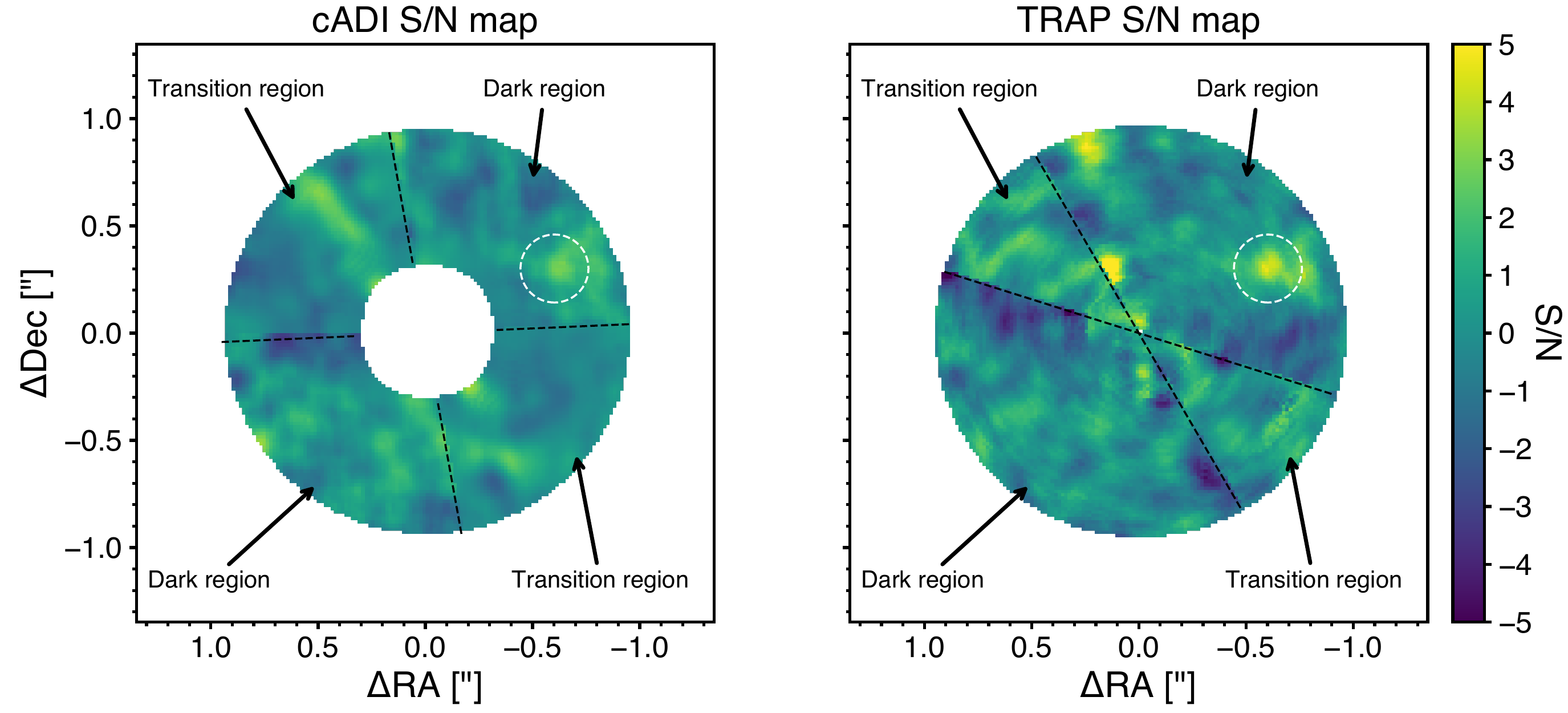}}
          \caption{Normalised S/N detection maps of cADI and TRAP for HR~2562. The dark region is where all the planet trajectories fall completely into the dark holes of the gvAPP180 PSF, while the transition region is where the planet moves across the dark and bright sides in the PSF. North is up and east is left. The dark region of the upper PSF is in the northwest and the dark region of the lower PSF is in the southeast. The transition region of cADI is enlarged to also include the severe speckle contamination near the boundary of the dark and bright sides of the PSF. The expected location of HR~2562~B is highlighted by the white circle. The bright and dark patterns near the edge of the dark regions in the S/N map of TRAP are results of the imperfect boundary between the bright and dark sides of the gvAPP180 PSF.
          }
 \label{Fig:gvAPP180_ADITRAP}
\end{figure*}

The S/N detection maps of cADI and TRAP are compared in Fig.~\ref{Fig:gvAPP180_ADITRAP}. TRAP has a stronger detection of HR~2562~B compared to cADI. 
The central excluded region of cADI is larger than the IWA due to cross-correlation used in the calculation. Table~\ref{Table_HR2562} lists the photometric measurement of HR~2562~B. Compared with cADI, TRAP increases the S/N by 54\%. Both measured contrasts from our cADI and TRAP agree with the contrast in \cite{Sutlieff2021} within 1$\sigma$, which is 3.05 $\pm$ 1.00 $\times 10^{-4}$. The contrast difference between our cADI and \cite{Sutlieff2021} is due to differences in implementation details, such as bad pixels correction, dark holes cropping and joining, planet position optimisation, and S/N calculation. 
But even if compared with the S/N measured by \cite{Sutlieff2021}, 3.04, TRAP still increases the S/N by 43\%.

\begin{table}[h!]
\centering
\setlength{\belowcaptionskip}{10pt}
\caption{Photometric measurement of HR~2562~B by cADI and TRAP.}
\begin{tabular}{l l l l}
\hline\hline
 Algorithm & Contrast ($10^{-4}$) & S/N \\
\hline
cADI & 2.63 $\pm$ 0.93 & 2.82 \\
TRAP & 2.91 $\pm$ 0.67 & 4.34 \\
Sutlieff2021 (cADI) & 3.05 $\pm$ 1.00 & 3.04 \\
\hline
\end{tabular}
\label{Table_HR2562}
\end{table}

\section{Discussion}
\label{sec:Discussion}
\subsection{HR~8799~e in the dgvAPP360 dataset}
In Sect.~\ref{sec:dgvAPP360_result} we reported the spectral results of HR~8799~cde with the custom algorithm and TRAP.
The innermost planet HR~8799~e has the lowest S/N and largest high-frequency noise in the contrast spectrum compared to the other two planets.
This is due to its proximity to the IWA of the coronagraph, where the noise pattern changes significantly from the stellar PSF core to the dark hole.
Moreover, only a small number of pixels are affected by the same underlying cause of systematic noise at this small separation, making it challenging to build either a temporal reference model or a spatial PSF model in this region.
As a result, both the custom algorithm and TRAP have difficulty detecting planet~e as effectively as the other two outer planets. This also explains why excluding reference pixels inside the inner working angle for TRAP when reducing the dark hole results in a higher S/N, since the noise patterns are quite different.

The parameters of the custom algorithm and TRAP were optimised for a specific wavelength channel, 3.81\,$\mu$m, where the transmission of the dgvAPP360 is high.
Optimising the custom algorithm for each channel to remove the wavelength dependent systematic noise may achieve smoother spectra.
TRAP can also produce better results if the parameters (e.g. the principal component number used to reconstruct the systematics and reference pixel selection) are optimised for each wavelength channel. 
However, based on our experience with optimised parameters in 3.81\,$\mu$m, the improvement of TRAP may not be obvious in this case because the results are insensitive to them once the parameters are close to the optimised parameters.
It is also computationally expensive to optimise parameters for each wavelength channel for a total number of 81 channels in both methods.
Since we wish to have a user-friendly and practical algorithm for vAPP coronagraphic data, optimising the parameters for each channel is beyond the scope of this work.

\subsection{Systematic noise in gvAPP180 PSFs}
\subsubsection{Confirmation by injection experiments on the Altair dataset}
In Sect.~\ref{sec:referenceTRAPgvAPP} we expected the dark holes of the upper and lower PSFs to share the most similar noise properties, since they are the only ones with starlight suppression. The joined dark holes were expected to be similar to a conventional coronagraphic PSF and thus share the most systematic causes. Therefore, we hypothesised that the joined dark holes would produce the best result.
However, the best result was found choosing reference pixels exclusively from the dark hole where the companion is located, running counter to our initial expectations.
To further investigate the noise behaviour in the four regions, we performed tests on the Altair dataset without any known companions. We injected a fake planet signal with a 5$\sigma$ contrast at the middle separation of the upper dark hole and retrieved it using reference pixels from the four regions, upper dark hole, upper bright side, lower dark hole, and lower bright side, respectively. More details on the reduction process for this dataset can be found in Appendix~\ref{apd}.
We did 5 injections for each choice, and Table~\ref{Table:Altair_injection} presents the averaged retrieved results.
Choosing reference pixels from the dark hole of the same PSF, the upper dark hole in this case, allows us to retrieve the signal with an uncertainty of within 1$\sigma$, as expected. It also has the smallest position deviation, which is within one pixel scale, indicating the uncertainty of the optimised planet position by TRAP. Choosing reference pixels from the bright side of the same PSF produces the second-highest S/N, $\sim$3.5, while choosing reference pixels from the complementary PSF has the worst performance, regardless from the dark hole or the bright side. 
This suggests that while the gvAPP180 coronagraph significantly suppresses starlight in the dark hole, there are still underlying noise causes that are shared by the dark hole and bright side of the same PSF.

\begin{table*}[h!]
\centering
\caption{5$\sigma$ injection to the upper dark region of the gvAPP180/Altair dataset and retrieved by choosing reference pixels from the upper dark hole, upper bright side, lower dark hole, and lower bright side. The results are averaged by five injections at a radial separation of 5$\lambda/D$ but at different angles. The uncertainty is the standard deviation of the five injections. The retrieved contrast deviation is relative to the injected contrast. The retrieved position deviation is absolute to the injected position.}
\begin{tabular}{l l l l}
\hline\hline
Reference pixels & S/N & Contrast deviation (\%) & Position deviation (mas) \\ 
\hline
Upper dark & 5.17 $\pm$ 0.73 & 15 $\pm$ 7 & 10 $\pm$ 3 \\
Upper bright & 3.54 $\pm$ 0.51 & 32 $\pm$ 22 & 17 $\pm$ 13 \\
Lower dark & 1.46 $\pm$ 3.24 & 53 $\pm$ 64 & 18 $\pm$ 11 \\
Lower bright & 1.42 $\pm$ 2.66 & 53 $\pm$ 80 & 24 $\pm$ 7 \\
\hline
\end{tabular}
\label{Table:Altair_injection}
\end{table*}

\subsubsection{Noise correlation between the two complementary PSFs}
The results of reference pixel experiments on both HR~2562 and Altair indicate that the noise varies significantly across the detector. In the two complementary PSFs, the noise caused by different sources is more pronounced than that resulting from the same sources.
The same causes of systematic noise shared by the two complementary PSFs are the dynamic aberrations and (quasi-)static optical aberrations between the telescope pupil and the coronagraph, such as the imperfect AO correction for the disturbance of the Earth atmosphere and non-common-path aberrations between the wavefront sensor and the coronagraph \citep[e.g.][]{Martinez2013}. After passing through the gvAPP coronagraph, the light path splits into two paths, which arrive at different two regions on the detector. This process introduces several factors that contribute to the systematic noise differences between the two complementary PSFs, including (quasi-)static optical aberrations after the coronagraph, instrument scattered lights, reflection ghosts, uncorrected response variance across the detector, and other artefacts \citep{Morzinski2015, Otten2017, Long2018}. Many of these factors are visible in the background-subtracted PSFs and affect different parts of the detector, exhibiting few homogeneous temporal variations. 
Therefore, TRAP cannot reconstruct the systematic noise for one PSF by choosing reference pixels from the complementary PSF.

The detection difference between cube A and cube B in Fig.~\ref{Fig:gvAPP180_ABSNRmap} is another piece of evidence for the noise variance across the detector. While the best reference pixel choice is the same for both cubes, the extracted signal from cube B is generally lower than that from cube A because the signal in cube B is more dispersed. 
The distance between A and B nod positions is even larger than that between the two complementary PSFs, resulting in less correlation between the noise of cubes A and and B. 
Visual inspection reveals that cube B is more affected by instrument scattered light and reflection ghosts on the detector, which explains why the extracted companion signal from cube B is lower than that from cube A. In fact, these artefacts also break the symmetry of the upper and lower PSFs, thereby degrading the performance of algorithms based on the symmetry of the gvAPP180 PSFs \citep{Sutlieff2021}.

\subsubsection{Noise correlation in the same PSF}
\begin{figure*}
\centering 
\resizebox{\hsize}{!}
          {\includegraphics[width=\textwidth]{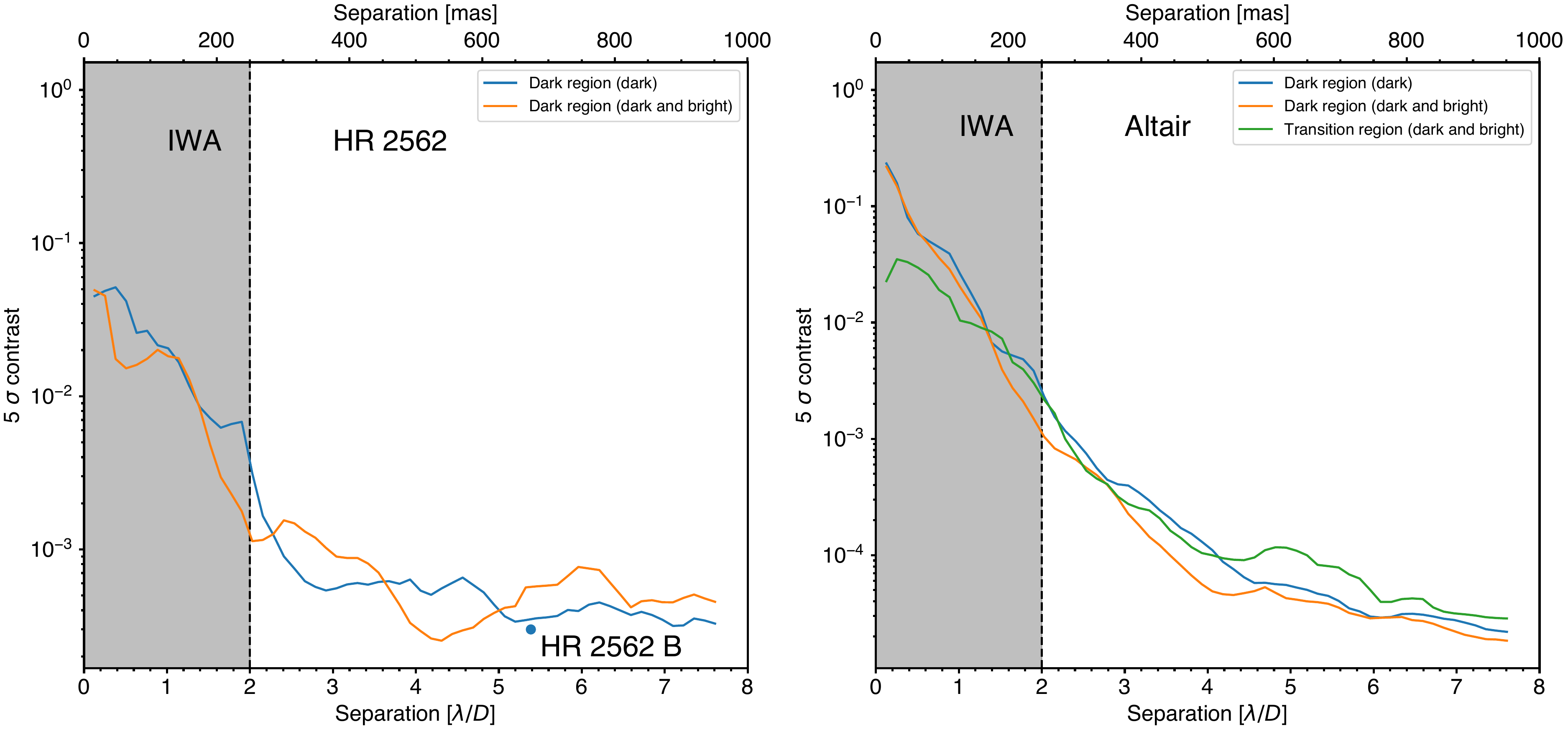}}
          \caption{5$\sigma$ contrast curves of the gvAPP180 datasets with TRAP. The contrast curve is calculated as five times the median of a three-pixel-wide annulus as a function of separation in the normalised uncertainty map. Left panel: the contrast curves of the upper dark region of cube A of the HR~2562 dataset reduced by choosing reference pixels exclusively from the dark hole (labelled as `dark') or both dark and bright sides (labelled as `dark and bright'). The detection significance of HR~2562~B is also marked in the figure. Right panel: the contrast curves of the dark region of the Altair dataset reduced by choosing reference pixels only from the dark hole or both dark and bright sides. The contrast curve of the transition region is also compared in the right panel, which is not much worse than that of the dark region.
          }
 \label{Fig:gvAPP180_contrast_curve}
\end{figure*}

We identified three sources of systematic noise that are still shared between the bright side and dark hole of the same PSF.
Firstly, the wind-driven halo around the central star \citep{Cantalloube2018,Madurowicz2019} keeps the noise at small angular separations still correlated on both dark and bright sides. 
Though we performed frame selections to remove frames strongly influenced by the stellar halo in the Altair dataset, there are still frames containing residual wind-driven halo extending to the dark hole. 
Secondly, two bright ghosts on the edge of the dark hole introduce noise from the bright side to the dark side. These wind-driven halo and ghosts are also discussed in \cite{Otten2017}, and are common problems for the MagAO/gvAPP180. Thirdly, although the gvAPP180 can discard the majority of starlight in the dark hole, a small remnant of the starlight and the diffraction pattern still exist in the dark hole. The raw contrast of the MagAO/gvAPP180 is $10^{-5}-10^{-4}$ in the dark hole \citep{Otten2017}, leaving the photon noise in the dark hole still higher than the noise in other empty regions of the detector. Some quasi-static wavefront aberrations induced by the instrument can also brighten the symmetric diffraction structure around the star and thus the corresponding part in the dark hole \citep{Otten2017}. Therefore, considerable part of the noise in the dark hole is still associated with the noise in the opposite bright side and thus shares similar temporal variations. These effects are particularly pronounced for bright primary stars, especially at small angular separations. So including pixels from the bright side into reference pixel selection can better model the systematic noise for bright stars. Whether one should include bright pixels for a faint star is a case dependent problem. As shown in the contrast curve of HR~2562 in Fig.~\ref{Fig:gvAPP180_contrast_curve}, including pixels from the bright side can improve contrast at certain separations, but there is no monotonic relationship. While the contrast curve of Altair in Fig.~\ref{Fig:gvAPP180_contrast_curve} shows that including pixels from the bright side indeed improves the contrast at small separations, the advantage recedes at larger separations.
When the primary star is faint, we suggest using the standard reference pixel setting (dark and bright pixels) for an initial blind search, and then only choosing dark reference pixels to test if it can achieve higher S/N for the companion if detected in the blind search. When the primary star is a bright star similar to Altair or the wind-driven halo is very strong during observation, using the standard reference pixel setting is recommended.

\subsection{Effective detection area of gvAPP180 PSFs}
The contrast curves of the transition region and dark region of Altair are also compared in the right panel of Fig.~\ref{Fig:gvAPP180_contrast_curve}. The contrast is not much worse than that of the dark region, partly because the transition region is independently reduced twice and combined from the upper and lower PSFs. It suggests that TRAP is able to detect a bright companion if it is unfortunately located in the transition region of gvAPP180. 

\begin{figure*}
\centering 
\resizebox{\hsize}{!}
          {\includegraphics[width=\textwidth]{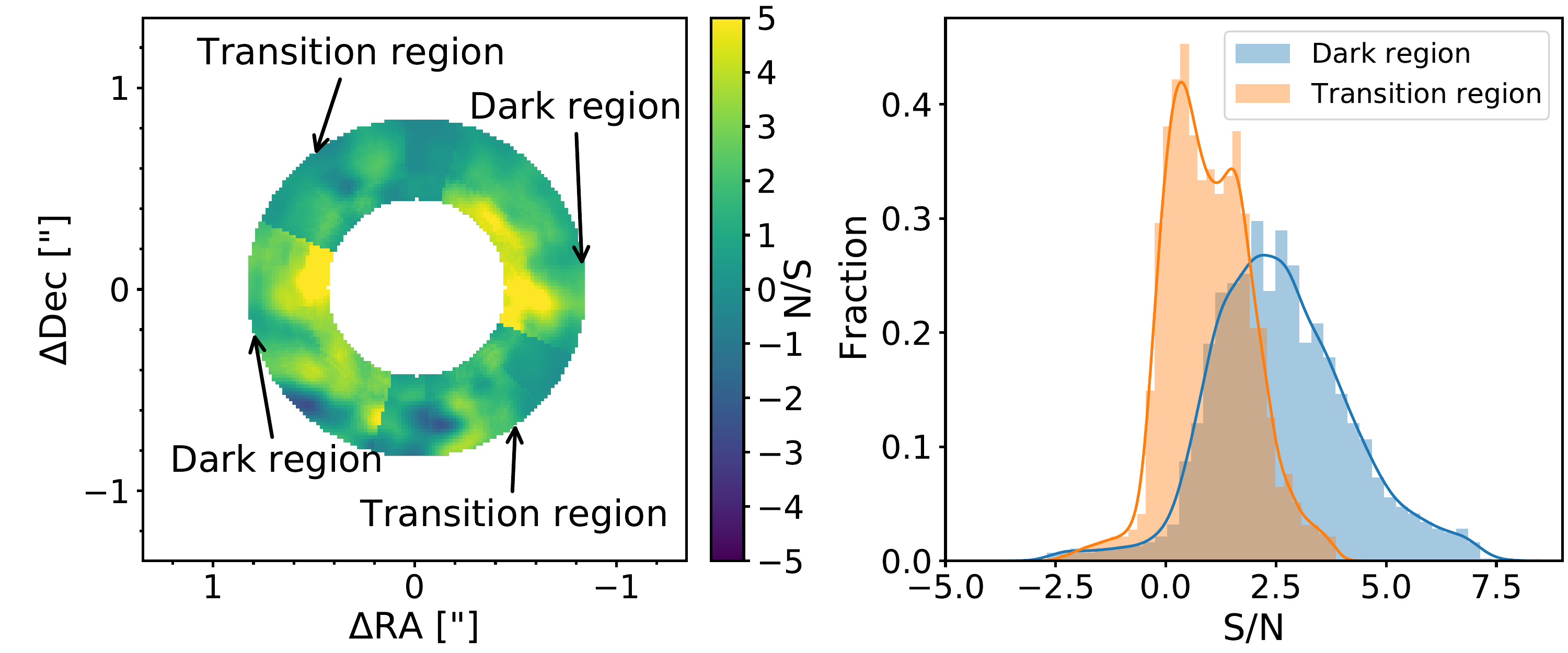}}
          \caption{S/N retrieved with cADI of 5$\sigma$ injection. The injected 5$\sigma$ signal at each position comes from the TRAP 5$\sigma$ map of the Altair dataset. Left panel: retrieved S/N map with cADI. The injected region is from 412 to 792 mas. The stronger detection at the inner part of the dark region is due to the higher false alarm probabilities caused by small sample statistics \citep{Mawet2014}. Only a small number of pixels are available for normalisation at small separations, and therefore the uncertainty is underestimated. Right panel: S/N histogram of the left panel. The blue line depicts the profile of the dark region with a median S/N of 2.5 and the orange line depicts of the profile of the transition region with a median S/N of 0.9.
          }
 \label{fig:gvAPP180_Altair_injection}
\end{figure*}

To compare the effective detection area between TRAP and cADI, we used the artificial planet injection method to compare the retrieved photometric measurement of the two algorithms. We took the 5$\sigma$ contrast map obtained by TRAP as the baseline and injected the 5$\sigma$ signals into the Altair dataset, covering 412--792\,mas. Each time we injected the signal at one position, and reduced the data by cADI and calculated the S/N of that position. Figure~\ref{fig:gvAPP180_Altair_injection} shows the measured S/N of all injected positions reduced by cADI. The retrieved S/N of cADI has a median of 2.5 in the dark region and is statistically less significant than 5$\sigma$. This is consistent with the results of HR~2562 that TRAP increases the S/N by 43\% compared to cADI. The median of the retrieved S/N of cADI in the transition region is only 0.9, much less significant than 5$\sigma$. It shows that TRAP also outperforms cADI in the transition region and the relative improvement is higher than that in the dark region.

Although it is preferable to ensure that the planet is located in the dark region of the gvAPP180 coronagraph for ADI observations to achieve higher S/N, this is not always possible for blind planet searches when the planet location is unknown beforehand. One solution is to rotate the gvAPP180 coronagraph to make the dark region cover an angle of $360^{\circ}$, which requires multiple observations of the same target. By improving post-processing algorithms such as TRAP, it is possible to detect relatively bright companions such as brown dwarfs, with an angle coverage of $360^{\circ}$ by a one-time pupil-tracking observation without rotating the gvAPP180 coronagraph.

\subsection{TRAP on dgvAPP360 and gvAPP180 datasets}
The dgvAPP360 and gvAPP180 coronagraphs produce different PSFs. We find that reference pixels from a symmetric annulus should be used for the dgvAPP360/HR~8799 dataset; for the gvAPP180/HR~2562 dataset only reference pixels in the upper dark annulus should be used; for the gvAPP180/Altair dataset reference pixels from symmetric annulus should be used. These reference pixel selections seem to differ between the dgvAPP360 and gvAPP180 datasets, but actually they follow the same selection strategy: choosing reference pixels from the same PSF as the supposed planet location. The dgvAPP360 only produces a single coronagrahic PSF, so reference pixels are chosen from this PSF; for the gvAPP180, the upper PSF should be used for the reduction of the upper dark hole and the lower PSF should be used for the reduction of the lower dark hole. 
A new gvAPP180 coronagraph has recently been commissioned on the Enhanced Resolution Imaging Spectrograph (ERIS) on the Very Large Telescope (VLT) and had its first on-sky results \citep{Kenworthy2018a,Boehle2021,Kravchenko2022}. If there are less artefacts, less static optical aberrations in the different path after the coronagraph and more homogeneous detector response in ERIS/NIX than MagAO/Clio2, reference pixels from the complementary PSF can also be used to reconstruct the systematic noise for the opposite PSF, which may increase the contrast at smaller separations where the available reference pixels are less. This is more promising for the future 39-m European Extremely Large Telescope (E-ELT), which will feature two gvAPP180 coronagraphs on the Mid-infrared ELT Imager and Spectrograph (METIS) \citep{Brandl2018} and Multi-AO Imaging Camera for Deep Observations (MICADO) \citep{Davies2016} instruments \citep{Kenworthy2018b,Doelman2021}.

\subsection{Outlook}
There is no perfect metric for comparing different algorithms as their underlying assumptions are different. Our comparison of TRAP, the custom speckle subtraction algorithm, and cADI is based on the photometric measurements of known companions to determine which algorithm can obtain the best contrast in the dark hole, rather than searching for potential companions in the field of view. Adopting metrics such as true positive rates, false positive rates, and F1-scores is helpful to further compare the detection maps of different algorithms \citep{Cantalloube2021}.

In Fig.~\ref{Fig:gvAPP180_ADITRAP} there are some dark and bright patterns near the border of the dark and transition regions in the S/N map of TRAP. These patterns result from the difficulty in modelling the systematic noise around the border of bright and dark sides. We used the half annulus in the dark hole as reference pixels, which is cut by a straight border between the dark and bright side, but the real border is more complex than a straight line.
This `half annulus' option is intended to best model the systematic noise in the middle of the dark hole where the planet is expected to be located. It is not an optimised way to search for companions near the transition region. How to optimally combine reference pixels from dark and bright sides to search for planets in gvAPP180 coronagraphic PSFs using TRAP is not covered in this work. One possible choice is to make the half annulus in Fig.~\ref{fig:gvAPP180_reference_pixels} move with the reduction region and always surround the reduction region in the centre instead of fixing the half annulus on the dark side. This choice may improve the detection on the edge of dark holes. For the IFS data, our implementation of algorithms is independent between different channels. Adding spectral differential imaging (SDI) to ADI-based algorithms can be applied to improve the detection \citep{Kiefer2021}.

As few techniques have been tested on vAPP coronagraphic data, even after adding TRAP tested in this work, we recommend exploring more algorithms, such as locally optimised combination of images (LOCI) \citep[e.g. ][]{Lafreniere2007, Wahhaj2015}, ANgular Differential OptiMal Exoplanet Detection Algorithm (ANDROMEDA) \citep{Cantalloube2015}, forward model matched filter (FMMF) \citep{Ruffio2017} and PACO \citep{Flasseur2018} to find the most robust algorithm for vAPP coronagraphic data. For gvAPP180 coronagraphic data, we suggest exploring more inverse problem techniques like TRAP. 
They may have better performance in the transition region than algorithms based on speckle subtraction because the inverse techniques track the planet signal in the data cube, estimate the maximum likelihood and some of them even do not require subtracting the reference PSF or derotating the science frames, which avoid introducing more artificial noise in post-processing procedures.
With more and more installation of vAPP coronagraphs \citep{Doelman2021}, testing different algorithms on vAPP coronagraphic data is important to prepare the community for the future observations and data reduction using these coronagraphs. As they suppress the star and planet PSF in the same way and preserves the central stellar PSF in per exposure, these coronagraphs are intrinsically suitable for variability monitoring of exoplanets and stellar companions by direct imaging \citep{Sutlieff2023}.

\section{Conclusions}
\label{sec:Conclusions}
By training a temporal model on non-local pixels, TRAP enables an investigation of
noise-time correlation between the dual PSFs of the gvAPP180 coronagraph.
We find that the systematic noise between the two PSFs of MagAO/gvAPP180 is quite different.
Unlike the conventional PSF, the symmetric systematic noise patterns between the two complementary dark holes are not significant and are overwhelmed by noise resulting from different sources. On the contrary, the bright side still shares a non-negligible amount of noise from the same underlying sources with the dark hole of the same PSF, even though the starlight is significantly suppressed in the dark hole. Thus we should only choose reference pixels from the same PSF when reducing the dark hole. If there is less systematic noise of different sources between the two PSFs, the opposite PSF can also be included to the temporal model, which will improve detections especially at small separations. This is promising for new generation of instruments, such as VLT/ERIS/gvAPP180, and is also instructive on the designing of the gvAPP180 coronagraphs on E-ELT.

In this work, we implemented a temporal algorithm, TRAP, on three vAPP coronagraphic datasets of two types of vAPP coronagraphs: the dgvAPP360 on LBT/ALES and the gvAPP180 on MagAO/Clio2. We compared the performance of TRAP with previously implemented algorithms that have the best S/N in companion detection, an ADI + PCA based custom speckle subtraction algorithm and classical ADI. For the dgvAPP360 dataset, TRAP achieves consistent results as the custom algorithm from \cite{Doelman2022}, with slightly higher S/N for the outer two planets HR~8799~c and d and a lower S/N for HR~8799~e. The contrast spectra of planet~e obtained by both algorithms present stronger high-frequency noise compared to the other planets.
For the gvAPP180 dataset, the contrasts retrieved by TRAP and cADI from \cite{Sutlieff2021} of HR~2562~B agree within one sigma, and TRAP improves the S/N by 43\%. TRAP also has the potential to detect relatively bright companions in the transition region where the planet trajectory spans the dark and bright sides in the gvAPP180 PSF. It makes the effective detection area of TRAP cover an angle of $360^{\circ}$ around the central star, while cADI can only have effective detections in the dark region.
The recommended choice for reference pixels of TRAP for dgvAPP360 datasets is the same as the standard setting, mainly a symmetric annulus at the same separation of the planet. For gvAPP180 datasets, when reducing the D-shaped PSF, it should be the annulus in the same PSF, either the half annulus in the dark hole or the full annulus including both the dark and bright sides.
We also encourage further testing of other post-processing algorithms on vAPP coronagraphic data, especially to achieve better detections near the IWA of the dgvAPP360 PSF and on the border between the dark hole and bright side of the gvAPP180 PSF.

\begin{acknowledgements}
We thank the referee for providing constructive comments that helped to improve the clarity of this paper. BJS is fully supported by the Netherlands Research School for Astronomy (NOVA).
JLB acknowledges funding from the European Research Council (ERC) under the European Union’s Horizon 2020 research and innovation program under grant agreement No 805445.
This paper includes data gathered with the 6.5 metre Magellan Telescopes located at Las Campanas Observatory, Chile, and the Large Binocular Telescope located in Arizona.
The LBT is an international collaboration among institutions in the United States, Italy and Germany. LBT Corporation partners are: The University of Arizona on behalf of the Arizona university system; Istituto Nazionale di Astrofisica, Italy; LBT Beteiligungsgesellschaft, Germany, representing the Max-Planck Society, the Astrophysical Institute Potsdam, and Heidelberg University; The Ohio State University, and The Research Corporation, on behalf of The University of Notre Dame, University of Minnesota, and University of Virginia. We thank all LBTI team members for their efforts that enabled this work. We gratefully acknowledge the use of Native land for our observations. LBT observations were conducted on the stolen land of the Ndee/Nnēē, Chiricahua, Mescalero and San Carlos Apache tribes.

This research made use of Astropy, a community-developed core Python package for Astronomy \citep{Astropy2013,Astropy2018}, SciPy \citep{Scipy2020}, Scikit-learn \citep{Scikit-learn2011}, NumPy \citep{Numpy2020} and Matplotlib, a Python library for publication quality graphics \citep{Matplotlib2007}.

\end{acknowledgements}

\bibliographystyle{aa} 
\bibliography{trap4vapp} 

\begin{thebibliography}{83}
\expandafter\ifx\csname natexlab\endcsname\relax\def\natexlab#1{#1}\fi

\bibitem[{{Apai} {et~al.}(2016){Apai}, {Kasper}, {Skemer}, {Hanson},
  {Lagrange}, {Biller}, {Bonnefoy}, {Buenzli}, \& {Vigan}}]{Apai2016}
{Apai}, D., {Kasper}, M., {Skemer}, A., {et~al.} 2016, \apj, 820, 40

\bibitem[{{Astropy Collaboration} {et~al.}(2018){Astropy Collaboration},
  {Price-Whelan}, {Sip{\H o}cz}, {G{\"u}nther}, {Lim}, {Crawford}, {Conseil},
  {Shupe}, {Craig}, {Dencheva}, {Ginsburg}, {VanderPlas}, {Bradley},
  {P{\'e}rez-Su{\'a}rez}, {de Val-Borro}, {Aldcroft}, {Cruz}, {Robitaille},
  {Tollerud}, {Ardelean}, {Babej}, {Bach}, {Bachetti}, {Bakanov}, {Bamford},
  {Barentsen}, {Barmby}, {Baumbach}, {Berry}, {Biscani}, {Boquien}, {Bostroem},
  {Bouma}, {Brammer}, {Bray}, {Breytenbach}, {Buddelmeijer}, {Burke},
  {Calderone}, {Cano Rodr{\'{\i}}guez}, {Cara}, {Cardoso}, {Cheedella},
  {Copin}, {Corrales}, {Crichton}, {D'Avella}, {Deil}, {Depagne}, {Dietrich},
  {Donath}, {Droettboom}, {Earl}, {Erben}, {Fabbro}, {Ferreira}, {Finethy},
  {Fox}, {Garrison}, {Gibbons}, {Goldstein}, {Gommers}, {Greco}, {Greenfield},
  {Groener}, {Grollier}, {Hagen}, {Hirst}, {Homeier}, {Horton}, {Hosseinzadeh},
  {Hu}, {Hunkeler}, {Ivezi{\'c}}, {Jain}, {Jenness}, {Kanarek}, {Kendrew},
  {Kern}, {Kerzendorf}, {Khvalko}, {King}, {Kirkby}, {Kulkarni}, {Kumar},
  {Lee}, {Lenz}, {Littlefair}, {Ma}, {Macleod}, {Mastropietro}, {McCully},
  {Montagnac}, {Morris}, {Mueller}, {Mumford}, {Muna}, {Murphy}, {Nelson},
  {Nguyen}, {Ninan}, {N{\"o}the}, {Ogaz}, {Oh}, {Parejko}, {Parley}, {Pascual},
  {Patil}, {Patil}, {Plunkett}, {Prochaska}, {Rastogi}, {Reddy Janga},
  {Sabater}, {Sakurikar}, {Seifert}, {Sherbert}, {Sherwood-Taylor}, {Shih},
  {Sick}, {Silbiger}, {Singanamalla}, {Singer}, {Sladen}, {Sooley},
  {Sornarajah}, {Streicher}, {Teuben}, {Thomas}, {Tremblay}, {Turner},
  {Terr{\'o}n}, {van Kerkwijk}, {de la Vega}, {Watkins}, {Weaver}, {Whitmore},
  {Woillez}, {Zabalza}, \& {Astropy Contributors}}]{Astropy2018}
{Astropy Collaboration}, {Price-Whelan}, A.~M., {Sip{\H o}cz}, B.~M., {et~al.}
  2018, \aj, 156, 123

\bibitem[{{Astropy Collaboration} {et~al.}(2013){Astropy Collaboration},
  {Robitaille}, {Tollerud}, {Greenfield}, {Droettboom}, {Bray}, {Aldcroft},
  {Davis}, {Ginsburg}, {Price-Whelan}, {Kerzendorf}, {Conley}, {Crighton},
  {Barbary}, {Muna}, {Ferguson}, {Grollier}, {Parikh}, {Nair}, {Unther},
  {Deil}, {Woillez}, {Conseil}, {Kramer}, {Turner}, {Singer}, {Fox}, {Weaver},
  {Zabalza}, {Edwards}, {Azalee Bostroem}, {Burke}, {Casey}, {Crawford},
  {Dencheva}, {Ely}, {Jenness}, {Labrie}, {Lim}, {Pierfederici}, {Pontzen},
  {Ptak}, {Refsdal}, {Servillat}, \& {Streicher}}]{Astropy2013}
{Astropy Collaboration}, {Robitaille}, T.~P., {Tollerud}, E.~J., {et~al.} 2013,
  \aap, 558, A33

\bibitem[{{Biller} {et~al.}(2021){Biller}, {Apai}, {Bonnefoy}, {Desidera},
  {Gratton}, {Kasper}, {Kenworthy}, {Lagrange}, {Lazzoni}, {Mesa}, {Vigan},
  {Wagner}, {Vos}, \& {Zurlo}}]{Biller2021}
{Biller}, B.~A., {Apai}, D., {Bonnefoy}, M., {et~al.} 2021, \mnras, 503, 743

\bibitem[{{Boehle} {et~al.}(2021){Boehle}, {Doelman}, {Konrad}, {Snik},
  {Glauser}, {Por}, {Warriner}, {Shi}, {Escuti}, {Kenworthy}, \&
  {Quanz}}]{Boehle2021}
{Boehle}, A., {Doelman}, D., {Konrad}, B.~S., {et~al.} 2021, Journal of
  Astronomical Telescopes, Instruments, and Systems, 7, 045001

\bibitem[{{Bohn} {et~al.}(2021){Bohn}, {Ginski}, {Kenworthy}, {Mamajek},
  {Pecaut}, {Mugrauer}, {Vogt}, {Adam}, {Meshkat}, {Reggiani}, \&
  {Snik}}]{Bohn2021}
{Bohn}, A.~J., {Ginski}, C., {Kenworthy}, M.~A., {et~al.} 2021, \aap, 648, A73

\bibitem[{{Bohn} {et~al.}(2020{\natexlab{a}}){Bohn}, {Kenworthy}, {Ginski},
  {Manara}, {Pecaut}, {de Boer}, {Keller}, {Mamajek}, {Meshkat}, {Reggiani},
  {Todorov}, \& {Snik}}]{Bohn2020a}
{Bohn}, A.~J., {Kenworthy}, M.~A., {Ginski}, C., {et~al.} 2020{\natexlab{a}},
  \mnras, 492, 431

\bibitem[{{Bohn} {et~al.}(2020{\natexlab{b}}){Bohn}, {Kenworthy}, {Ginski},
  {Rieder}, {Mamajek}, {Meshkat}, {Pecaut}, {Reggiani}, {de Boer}, {Keller},
  {Snik}, \& {Southworth}}]{Bohn2020b}
{Bohn}, A.~J., {Kenworthy}, M.~A., {Ginski}, C., {et~al.} 2020{\natexlab{b}},
  \apjl, 898, L16

\bibitem[{{Bonavita} {et~al.}(2022){Bonavita}, {Fontanive}, {Gratton},
  {Mu{\v{z}}i{\'c}}, {Desidera}, {Mesa}, {Biller}, {Scholz}, {Sozzetti}, \&
  {Squicciarini}}]{Bonavita2022}
{Bonavita}, M., {Fontanive}, C., {Gratton}, R., {et~al.} 2022, \mnras, 513,
  5588

\bibitem[{{Bonnefoy} {et~al.}(2016){Bonnefoy}, {Zurlo}, {Baudino}, {Lucas},
  {Mesa}, {Maire}, {Vigan}, {Galicher}, {Homeier}, {Marocco}, {Gratton},
  {Chauvin}, {Allard}, {Desidera}, {Kasper}, {Moutou}, {Lagrange}, {Antichi},
  {Baruffolo}, {Baudrand}, {Beuzit}, {Boccaletti}, {Cantalloube}, {Carbillet},
  {Charton}, {Claudi}, {Costille}, {Dohlen}, {Dominik}, {Fantinel},
  {Feautrier}, {Feldt}, {Fusco}, {Gigan}, {Girard}, {Gluck}, {Gry}, {Henning},
  {Janson}, {Langlois}, {Madec}, {Magnard}, {Maurel}, {Mawet}, {Meyer},
  {Milli}, {Moeller-Nilsson}, {Mouillet}, {Pavlov}, {Perret}, {Pujet}, {Quanz},
  {Rochat}, {Rousset}, {Roux}, {Salasnich}, {Salter}, {Sauvage}, {Schmid},
  {Sevin}, {Soenke}, {Stadler}, {Turatto}, {Udry}, {Vakili}, {Wahhaj}, \&
  {Wildi}}]{Bonnefoy2016}
{Bonnefoy}, M., {Zurlo}, A., {Baudino}, J.~L., {et~al.} 2016, \aap, 587, A58

\bibitem[{{Bonse} {et~al.}(2018){Bonse}, {Quanz}, \& {Amara}}]{Bonse2018}
{Bonse}, M.~J., {Quanz}, S.~P., \& {Amara}, A. 2018, arXiv e-prints,
  arXiv:1804.05063

\bibitem[{{Brandl} {et~al.}(2018){Brandl}, {Absil}, {Ag{\'o}cs}, {Baccichet},
  {Bertram}, {Bettonvil}, {van Boekel}, {Burtscher}, {van Dishoeck}, {Feldt},
  {Garcia}, {Glasse}, {Glauser}, {G{\"u}del}, {Haupt}, {Kenworthy}, {Labadie},
  {Laun}, {Lesman}, {Pantin}, {Quanz}, {Snellen}, {Siebenmorgen}, \& {van
  Winckel}}]{Brandl2018}
{Brandl}, B.~R., {Absil}, O., {Ag{\'o}cs}, T., {et~al.} 2018, in Society of
  Photo-Optical Instrumentation Engineers (SPIE) Conference Series, Vol. 10702,
  Ground-based and Airborne Instrumentation for Astronomy VII, ed. C.~J.
  {Evans}, L.~{Simard}, \& H.~{Takami}, 107021U

\bibitem[{{Brandt} {et~al.}(2021){Brandt}, {Brandt}, {Dupuy}, {Michalik}, \&
  {Marleau}}]{Brandt2021}
{Brandt}, G.~M., {Brandt}, T.~D., {Dupuy}, T.~J., {Michalik}, D., \& {Marleau},
  G.-D. 2021, \apjl, 915, L16

\bibitem[{{Cantalloube} {et~al.}(2021){Cantalloube}, {Gomez-Gonzalez}, {Absil},
  {Cantero}, {Bacher}, {Bonse}, {Bottom}, {Dahlqvist}, {Desgrange}, {Flasseur},
  {Fuhrmann}, {Henning}, {Jensen-Clem}, {Kenworthy}, {Mawet}, {Mesa},
  {Meshkat}, {Mouillet}, {Mueller}, {Nasedkin}, {Pairet}, {Pierard}, {Ruffio},
  {Samland}, {Stone}, \& {Van Droogenbroeck}}]{Cantalloube2021}
{Cantalloube}, F., {Gomez-Gonzalez}, C., {Absil}, O., {et~al.} 2021, arXiv
  e-prints, arXiv:2101.05080

\bibitem[{{Cantalloube} {et~al.}(2015){Cantalloube}, {Mouillet}, {Mugnier},
  {Milli}, {Absil}, {Gomez Gonzalez}, {Chauvin}, {Beuzit}, \&
  {Cornia}}]{Cantalloube2015}
{Cantalloube}, F., {Mouillet}, D., {Mugnier}, L.~M., {et~al.} 2015, \aap, 582,
  A89

\bibitem[{{Cantalloube} {et~al.}(2018){Cantalloube}, {Por}, {Dohlen},
  {Sauvage}, {Vigan}, {Kasper}, {Bharmal}, {Henning}, {Brandner}, {Milli},
  {Correia}, \& {Fusco}}]{Cantalloube2018}
{Cantalloube}, F., {Por}, E.~H., {Dohlen}, K., {et~al.} 2018, \aap, 620, L10

\bibitem[{{Chauvin} {et~al.}(2017){Chauvin}, {Desidera}, {Lagrange}, {Vigan},
  {Gratton}, {Langlois}, {Bonnefoy}, {Beuzit}, {Feldt}, {Mouillet}, {Meyer},
  {Cheetham}, {Biller}, {Boccaletti}, {D'Orazi}, {Galicher}, {Hagelberg},
  {Maire}, {Mesa}, {Olofsson}, {Samland}, {Schmidt}, {Sissa}, {Bonavita},
  {Charnay}, {Cudel}, {Daemgen}, {Delorme}, {Janin-Potiron}, {Janson},
  {Keppler}, {Le Coroller}, {Ligi}, {Marleau}, {Messina}, {Molli{\`e}re},
  {Mordasini}, {M{\"u}ller}, {Peretti}, {Perrot}, {Rodet}, {Rouan}, {Zurlo},
  {Dominik}, {Henning}, {Menard}, {Schmid}, {Turatto}, {Udry}, {Vakili}, {Abe},
  {Antichi}, {Baruffolo}, {Baudoz}, {Baudrand}, {Blanchard}, {Bazzon}, {Buey},
  {Carbillet}, {Carle}, {Charton}, {Cascone}, {Claudi}, {Costille}, {Deboulbe},
  {De Caprio}, {Dohlen}, {Fantinel}, {Feautrier}, {Fusco}, {Gigan}, {Giro},
  {Gisler}, {Gluck}, {Hubin}, {Hugot}, {Jaquet}, {Kasper}, {Madec}, {Magnard},
  {Martinez}, {Maurel}, {Le Mignant}, {M{\"o}ller-Nilsson}, {Llored}, {Moulin},
  {Orign{\'e}}, {Pavlov}, {Perret}, {Petit}, {Pragt}, {Puget}, {Rabou},
  {Ramos}, {Rigal}, {Rochat}, {Roelfsema}, {Rousset}, {Roux}, {Salasnich},
  {Sauvage}, {Sevin}, {Soenke}, {Stadler}, {Suarez}, {Weber}, {Wildi},
  {Antoniucci}, {Augereau}, {Baudino}, {Brandner}, {Engler}, {Girard}, {Gry},
  {Kral}, {Kopytova}, {Lagadec}, {Milli}, {Moutou}, {Schlieder},
  {Szul{\'a}gyi}, {Thalmann}, \& {Wahhaj}}]{Chauvin2017}
{Chauvin}, G., {Desidera}, S., {Lagrange}, A.~M., {et~al.} 2017, \aap, 605, L9

\bibitem[{{Close} {et~al.}(2010){Close}, {Gasho}, {Kopon}, {Males}, {Follette},
  {Brutlag}, {Uomoto}, \& {Hare}}]{Close2010}
{Close}, L.~M., {Gasho}, V., {Kopon}, D., {et~al.} 2010, in Society of
  Photo-Optical Instrumentation Engineers (SPIE) Conference Series, Vol. 7736,
  Adaptive Optics Systems II, ed. B.~L. {Ellerbroek}, M.~{Hart}, N.~{Hubin}, \&
  P.~L. {Wizinowich}, 773605

\bibitem[{{Currie} {et~al.}(2014){Currie}, {Burrows}, {Girard}, {Cloutier},
  {Fukagawa}, {Sorahana}, {Kuchner}, {Kenyon}, {Madhusudhan}, {Itoh},
  {Jayawardhana}, {Matsumura}, \& {Pyo}}]{Currie2014}
{Currie}, T., {Burrows}, A., {Girard}, J.~H., {et~al.} 2014, \apj, 795, 133

\bibitem[{{Davies} {et~al.}(2016){Davies}, {Schubert}, {Hartl}, {Alves},
  {Cl{\'e}net}, {Lang-Bardl}, {Nicklas}, {Pott}, {Ragazzoni}, {Tolstoy},
  {Agocs}, {Anwand-Heerwart}, {Barboza}, {Baudoz}, {Bender}, {Bizenberger},
  {Boccaletti}, {Boland}, {Bonifacio}, {Briegel}, {Buey}, {Chapron}, {Cohen},
  {Czoske}, {Dreizler}, {Falomo}, {Feautrier}, {F{\"o}rster Schreiber},
  {Gendron}, {Genzel}, {Gl{\"u}ck}, {Gratadour}, {Greimel}, {Grupp},
  {H{\"a}user}, {Haug}, {Hennawi}, {Hess}, {H{\"o}rmann}, {Hofferbert}, {Hopp},
  {Hubert}, {Ives}, {Kausch}, {Kerber}, {Kravcar}, {Kuijken}, {Lang-Bardl},
  {Leitzinger}, {Leschinski}, {Massari}, {Mei}, {Merlin}, {Mohr}, {Monna},
  {M{\"u}ller}, {Navarro}, {Plattner}, {Przybilla}, {Ramlau}, {Ramsay},
  {Ratzka}, {Rhode}, {Richter}, {Rix}, {Rodeghiero}, {Rohloff}, {Rousset},
  {Ruddenklau}, {Schaffenroth}, {Schlichter}, {Sevin}, {Stuik}, {Sturm},
  {Thomas}, {Tromp}, {Turatto}, {Verdoes-Kleijn}, {Vidal}, {Wagner}, {Wegner},
  {Zeilinger}, {Ziegler}, \& {Zins}}]{Davies2016}
{Davies}, R., {Schubert}, J., {Hartl}, M., {et~al.} 2016, in Society of
  Photo-Optical Instrumentation Engineers (SPIE) Conference Series, Vol. 9908,
  Ground-based and Airborne Instrumentation for Astronomy VI, ed. C.~J.
  {Evans}, L.~{Simard}, \& H.~{Takami}, 99081Z

\bibitem[{{Desidera} {et~al.}(2021){Desidera}, {Chauvin}, {Bonavita},
  {Messina}, {LeCoroller}, {Schmidt}, {Gratton}, {Lazzoni}, {Meyer},
  {Schlieder}, {Cheetham}, {Hagelberg}, {Bonnefoy}, {Feldt}, {Lagrange},
  {Langlois}, {Vigan}, {Tan}, {Hambsch}, {Millward}, {Alcal{\'a}}, {Benatti},
  {Brandner}, {Carson}, {Covino}, {Delorme}, {D'Orazi}, {Janson}, {Rigliaco},
  {Beuzit}, {Biller}, {Boccaletti}, {Dominik}, {Cantalloube}, {Fontanive},
  {Galicher}, {Henning}, {Lagadec}, {Ligi}, {Maire}, {Menard}, {Mesa},
  {M{\"u}ller}, {Samland}, {Schmid}, {Sissa}, {Turatto}, {Udry}, {Zurlo},
  {Asensio-Torres}, {Kopytova}, {Rickman}, {Abe}, {Antichi}, {Baruffolo},
  {Baudoz}, {Baudrand}, {Blanchard}, {Bazzon}, {Buey}, {Carbillet}, {Carle},
  {Charton}, {Cascone}, {Claudi}, {Costille}, {Deboulb{\'e}}, {De Caprio},
  {Dohlen}, {Fantinel}, {Feautrier}, {Fusco}, {Gigan}, {Giro}, {Gisler},
  {Gluck}, {Hubin}, {Hugot}, {Jaquet}, {Kasper}, {Madec}, {Magnard},
  {Martinez}, {Maurel}, {Le Mignant}, {M{\"o}ller-Nilsson}, {Llored}, {Moulin},
  {Orign{\'e}}, {Pavlov}, {Perret}, {Petit}, {Pragt}, {Puget}, {Rabou},
  {Ramos}, {Rigal}, {Rochat}, {Roelfsema}, {Rousset}, {Roux}, {Salasnich},
  {Sauvage}, {Sevin}, {Soenke}, {Stadler}, {Suarez}, {Weber}, \&
  {Wildi}}]{Desidera2021}
{Desidera}, S., {Chauvin}, G., {Bonavita}, M., {et~al.} 2021, \aap, 651, A70

\bibitem[{{Doelman} {et~al.}(2020){Doelman}, {Por}, {Ruane}, {Escuti}, \&
  {Snik}}]{Doelman2020}
{Doelman}, D.~S., {Por}, E.~H., {Ruane}, G., {Escuti}, M.~J., \& {Snik}, F.
  2020, \pasp, 132, 045002

\bibitem[{{Doelman} {et~al.}(2021){Doelman}, {Snik}, {Por}, {Bos}, {Otten},
  {Kenworthy}, {Haffert}, {Wilby}, {Bohn}, {Sutlieff}, {Miller}, {Ouellet}, {de
  Boer}, {Keller}, {Escuti}, {Shi}, {Warriner}, {Hornburg}, {Birkby}, {Males},
  {Morzinski}, {Close}, {Codona}, {Long}, {Schatz}, {Lumbres}, {Rodack}, {Van
  Gorkom}, {Hedglen}, {Guyon}, {Lozi}, {Groff}, {Chilcote}, {Jovanovic},
  {Thibault}, {de Jonge}, {Allain}, {Vall{\'e}e}, {Patel}, {C{\^o}t{\'e}},
  {Marois}, {Hinz}, {Stone}, {Skemer}, {Briesemeister}, {Boehle}, {Glauser},
  {Taylor}, {Baudoz}, {Huby}, {Absil}, {Carlomagno}, \&
  {Delacroix}}]{Doelman2021}
{Doelman}, D.~S., {Snik}, F., {Por}, E.~H., {et~al.} 2021, \ao, 60, D52

\bibitem[{{Doelman} {et~al.}(2022){Doelman}, {Stone}, {Briesemeister},
  {Skemer}, {Barman}, {Brock}, {Hinz}, {Bohn}, {Kenworthy}, {Haffert}, {Snik},
  {Ertel}, {Leisenring}, {Woodward}, \& {Skrutskie}}]{Doelman2022}
{Doelman}, D.~S., {Stone}, J.~M., {Briesemeister}, Z.~W., {et~al.} 2022, \aj,
  163, 217

\bibitem[{{Flasseur} {et~al.}(2018){Flasseur}, {Denis}, {Thi{\'e}baut}, \&
  {Langlois}}]{Flasseur2018}
{Flasseur}, O., {Denis}, L., {Thi{\'e}baut}, {\'E}., \& {Langlois}, M. 2018,
  \aap, 618, A138

\bibitem[{{Gebhard} {et~al.}(2022){Gebhard}, {Bonse}, {Quanz}, \&
  {Sch{\"o}lkopf}}]{Gebhard2022}
{Gebhard}, T.~D., {Bonse}, M.~J., {Quanz}, S.~P., \& {Sch{\"o}lkopf}, B. 2022,
  \aap, 666, A9

\bibitem[{{Go{\'z}dziewski} \& {Migaszewski}(2020)}]{dziewski2020}
{Go{\'z}dziewski}, K. \& {Migaszewski}, C. 2020, \apjl, 902, L40

\bibitem[{{Gray} {et~al.}(2006){Gray}, {Corbally}, {Garrison}, {McFadden},
  {Bubar}, {McGahee}, {O'Donoghue}, \& {Knox}}]{Gray2006}
{Gray}, R.~O., {Corbally}, C.~J., {Garrison}, R.~F., {et~al.} 2006, \aj, 132,
  161

\bibitem[{{Gray} {et~al.}(2003){Gray}, {Corbally}, {Garrison}, {McFadden}, \&
  {Robinson}}]{Gray2003}
{Gray}, R.~O., {Corbally}, C.~J., {Garrison}, R.~F., {McFadden}, M.~T., \&
  {Robinson}, P.~E. 2003, \aj, 126, 2048

\bibitem[{{Gray} \& {Kaye}(1999)}]{Gray1999}
{Gray}, R.~O. \& {Kaye}, A.~B. 1999, \aj, 118, 2993

\bibitem[{Harris {et~al.}(2020)Harris, Millman, van~der Walt, Gommers,
  Virtanen, Cournapeau, Wieser, Taylor, Berg, Smith, Kern, Picus, Hoyer, van
  Kerkwijk, Brett, Haldane, del R{'{\i}}o, Wiebe, Peterson,
  G{'{e}}rard-Marchant, Sheppard, Reddy, Weckesser, Abbasi, Gohlke, \&
  Oliphant}]{Numpy2020}
Harris, C.~R., Millman, K.~J., van~der Walt, S.~J., {et~al.} 2020, Nature, 585,
  357

\bibitem[{Hunter(2007)}]{Matplotlib2007}
Hunter, J.~D. 2007, Computing In Science \& Engineering, 9, 90

\bibitem[{{Janson} {et~al.}(2019){Janson}, {Asensio-Torres}, {Andr{\'e}},
  {Bonnefoy}, {Delorme}, {Reffert}, {Desidera}, {Langlois}, {Chauvin},
  {Gratton}, {Bohn}, {Eriksson}, {Marleau}, {Mamajek}, {Vigan}, \&
  {Carson}}]{Janson2019}
{Janson}, M., {Asensio-Torres}, R., {Andr{\'e}}, D., {et~al.} 2019, \aap, 626,
  A99

\bibitem[{{Kenworthy} {et~al.}(2018{\natexlab{a}}){Kenworthy}, {Absil},
  {Carlomagno}, {Ag{\'o}cs}, {Por}, {Bos}, {Brandl}, \&
  {Snik}}]{Kenworthy2018b}
{Kenworthy}, M.~A., {Absil}, O., {Carlomagno}, B., {et~al.} 2018{\natexlab{a}},
  in Society of Photo-Optical Instrumentation Engineers (SPIE) Conference
  Series, Vol. 10702, Ground-based and Airborne Instrumentation for Astronomy
  VII, ed. C.~J. {Evans}, L.~{Simard}, \& H.~{Takami}, 10702A3

\bibitem[{{Kenworthy} {et~al.}(2007){Kenworthy}, {Codona}, {Hinz}, {Angel},
  {Heinze}, \& {Sivanandam}}]{Kenworthy2007}
{Kenworthy}, M.~A., {Codona}, J.~L., {Hinz}, P.~M., {et~al.} 2007, \apj, 660,
  762

\bibitem[{{Kenworthy} {et~al.}(2018{\natexlab{b}}){Kenworthy}, {Snik},
  {Keller}, {Doelman}, {Por}, {Absil}, {Carlomagno}, {Karlsson}, {Huby},
  {Glauser}, {Quanz}, \& {Taylor}}]{Kenworthy2018a}
{Kenworthy}, M.~A., {Snik}, F., {Keller}, C.~U., {et~al.} 2018{\natexlab{b}},
  in Society of Photo-Optical Instrumentation Engineers (SPIE) Conference
  Series, Vol. 10702, Ground-based and Airborne Instrumentation for Astronomy
  VII, ed. C.~J. {Evans}, L.~{Simard}, \& H.~{Takami}, 1070246

\bibitem[{{Kiefer} {et~al.}(2021){Kiefer}, {Bohn}, {Quanz}, {Kenworthy}, \&
  {Stolker}}]{Kiefer2021}
{Kiefer}, S., {Bohn}, A.~J., {Quanz}, S.~P., {Kenworthy}, M., \& {Stolker}, T.
  2021, \aap, 652, A33

\bibitem[{{Konopacky} {et~al.}(2016){Konopacky}, {Rameau}, {Duch{\^e}ne},
  {Filippazzo}, {Giorla Godfrey}, {Marois}, {Nielsen}, {Pueyo}, {Rafikov},
  {Rice}, {Wang}, {Ammons}, {Bailey}, {Barman}, {Bulger}, {Bruzzone},
  {Chilcote}, {Cotten}, {Dawson}, {De Rosa}, {Doyon}, {Esposito}, {Fitzgerald},
  {Follette}, {Goodsell}, {Graham}, {Greenbaum}, {Hibon}, {Hung}, {Ingraham},
  {Kalas}, {Lafreni{\`e}re}, {Larkin}, {Macintosh}, {Maire}, {Marchis},
  {Marley}, {Matthews}, {Metchev}, {Millar-Blanchaer}, {Oppenheimer}, {Palmer},
  {Patience}, {Perrin}, {Poyneer}, {Rajan}, {Rantakyr{\"o}}, {Savransky},
  {Schneider}, {Sivaramakrishnan}, {Song}, {Soummer}, {Thomas}, {Wallace},
  {Ward-Duong}, {Wiktorowicz}, \& {Wolff}}]{Konopacky2016}
{Konopacky}, Q.~M., {Rameau}, J., {Duch{\^e}ne}, G., {et~al.} 2016, \apjl, 829,
  L4

\bibitem[{{Kravchenko} {et~al.}(2022){Kravchenko}, {Dallilar}, {Absil},
  {Berbel}, {Baruffolo}, {Bonse}, {Buron}, {Cao}, {Cortes}, {Dannert},
  {Davies}, {De Rosa}, {Deysenroth}, {Doelman}, {Eisenhauer}, {Esposito},
  {Feuchtgruber}, {F{\"o}rster Schreiber}, {Gao}, {Gemperlein}, {Genzel},
  {Gillessen}, {Ginski}, {Glauser}, {Glindemann}, {Grani}, {Haguenauer},
  {Hartwig}, {Hayoz}, {Heida}, {Kenworthy}, {Kolb}, {Kuntschner}, {Lutz},
  {Liu}, {MacIntosh}, {Marsset}, {Orban de Xivry}, {{\"O}zdemir}, {Puglisi},
  {Quanz}, {Rau}, {Riccardi}, {Schuppe}, {Snik}, {Sturm}, {Tacconi}, {Taylor},
  \& {Wiezorrek}}]{Kravchenko2022}
{Kravchenko}, K., {Dallilar}, Y., {Absil}, O., {et~al.} 2022, in Society of
  Photo-Optical Instrumentation Engineers (SPIE) Conference Series, Vol. 12184,
  Ground-based and Airborne Instrumentation for Astronomy IX, ed. C.~J.
  {Evans}, J.~J. {Bryant}, \& K.~{Motohara}, 121845M

\bibitem[{{Lafreni{\`e}re} {et~al.}(2007){Lafreni{\`e}re}, {Marois}, {Doyon},
  {Nadeau}, \& {Artigau}}]{Lafreniere2007}
{Lafreni{\`e}re}, D., {Marois}, C., {Doyon}, R., {Nadeau}, D., \& {Artigau},
  {\'E}. 2007, \apj, 660, 770

\bibitem[{{Lagrange} {et~al.}(2010){Lagrange}, {Bonnefoy}, {Chauvin}, {Apai},
  {Ehrenreich}, {Boccaletti}, {Gratadour}, {Rouan}, {Mouillet}, {Lacour}, \&
  {Kasper}}]{Lagrange2010}
{Lagrange}, A.~M., {Bonnefoy}, M., {Chauvin}, G., {et~al.} 2010, Science, 329,
  57

\bibitem[{{Langlois} {et~al.}(2021){Langlois}, {Gratton}, {Lagrange},
  {Delorme}, {Boccaletti}, {Bonnefoy}, {Maire}, {Mesa}, {Chauvin}, {Desidera},
  {Vigan}, {Cheetham}, {Hagelberg}, {Feldt}, {Meyer}, {Rubini}, {Le Coroller},
  {Cantalloube}, {Biller}, {Bonavita}, {Bhowmik}, {Brandner}, {Daemgen},
  {D'Orazi}, {Flasseur}, {Fontanive}, {Galicher}, {Girard}, {Janin-Potiron},
  {Janson}, {Keppler}, {Kopytova}, {Lagadec}, {Lannier}, {Lazzoni}, {Ligi},
  {Meunier}, {Perreti}, {Perrot}, {Rodet}, {Romero}, {Rouan}, {Samland},
  {Salter}, {Sissa}, {Schmidt}, {Zurlo}, {Mouillet}, {Denis}, {Thi{\'e}baut},
  {Milli}, {Wahhaj}, {Beuzit}, {Dominik}, {Henning}, {M{\'e}nard},
  {M{\"u}ller}, {Schmid}, {Turatto}, {Udry}, {Abe}, {Antichi}, {Allard},
  {Baruffolo}, {Baudoz}, {Baudrand}, {Bazzon}, {Blanchard}, {Carbillet},
  {Carle}, {Cascone}, {Charton}, {Claudi}, {Costille}, {De Caprio},
  {Delboulb{\'e}}, {Dohlen}, {Fantinel}, {Feautrier}, {Fusco}, {Gigan}, {Giro},
  {Gisler}, {Gluck}, {Gry}, {Hubin}, {Hugot}, {Jaquet}, {Kasper}, {Le Mignant},
  {Llored}, {Madec}, {Magnard}, {Martinez}, {Maurel}, {Messina},
  {M{\"o}ller-Nilsson}, {Mugnier}, {Moulin}, {Orign{\'e}}, {Pavlov}, {Perret},
  {Petit}, {Pragt}, {Puget}, {Rabou}, {Ramos}, {Rigal}, {Rochat}, {Roelfsema},
  {Rousset}, {Roux}, {Salasnich}, {Sauvage}, {Sevin}, {Soenke}, {Stadler},
  {Suarez}, {Weber}, {Wildi}, \& {Rickman}}]{Langlois2021}
{Langlois}, M., {Gratton}, R., {Lagrange}, A.~M., {et~al.} 2021, \aap, 651, A71

\bibitem[{{Lewis} {et~al.}(2023){Lewis}, {Fitzgerald}, {Dodkins}, {Davis}, \&
  {Lin}}]{Lewis2023}
{Lewis}, B., {Fitzgerald}, M.~P., {Dodkins}, R.~H., {Davis}, K.~K., \& {Lin},
  J. 2023, \aj, 165, 59

\bibitem[{{Long} {et~al.}(2023){Long}, {Males}, {Haffert}, {Pearce}, {Marley},
  {Morzinski}, {Close}, {Otten}, {Snik}, {Kenworthy}, {Keller}, {Hinz},
  {Monnier}, {Weinberger}, \& {Tolls}}]{Long2023}
{Long}, J.~D., {Males}, J.~R., {Haffert}, S.~Y., {et~al.} 2023, arXiv e-prints,
  arXiv:2303.05559

\bibitem[{{Long} {et~al.}(2018){Long}, {Males}, {Morzinski}, {Close}, {Snik},
  {Kenworthy}, {Otten}, {Monnier}, {Tolls}, \& {Weinberger}}]{Long2018}
{Long}, J.~D., {Males}, J.~R., {Morzinski}, K.~M., {et~al.} 2018, in Society of
  Photo-Optical Instrumentation Engineers (SPIE) Conference Series, Vol. 10703,
  Adaptive Optics Systems VI, ed. L.~M. {Close}, L.~{Schreiber}, \&
  D.~{Schmidt}, 107032T

\bibitem[{{Macintosh} {et~al.}(2015){Macintosh}, {Graham}, {Barman}, {De Rosa},
  {Konopacky}, {Marley}, {Marois}, {Nielsen}, {Pueyo}, {Rajan}, {Rameau},
  {Saumon}, {Wang}, {Patience}, {Ammons}, {Arriaga}, {Artigau}, {Beckwith},
  {Brewster}, {Bruzzone}, {Bulger}, {Burningham}, {Burrows}, {Chen}, {Chiang},
  {Chilcote}, {Dawson}, {Dong}, {Doyon}, {Draper}, {Duch{\^e}ne}, {Esposito},
  {Fabrycky}, {Fitzgerald}, {Follette}, {Fortney}, {Gerard}, {Goodsell},
  {Greenbaum}, {Hibon}, {Hinkley}, {Cotten}, {Hung}, {Ingraham},
  {Johnson-Groh}, {Kalas}, {Lafreniere}, {Larkin}, {Lee}, {Line}, {Long},
  {Maire}, {Marchis}, {Matthews}, {Max}, {Metchev}, {Millar-Blanchaer},
  {Mittal}, {Morley}, {Morzinski}, {Murray-Clay}, {Oppenheimer}, {Palmer},
  {Patel}, {Perrin}, {Poyneer}, {Rafikov}, {Rantakyr{\"o}}, {Rice}, {Rojo},
  {Rudy}, {Ruffio}, {Ruiz}, {Sadakuni}, {Saddlemyer}, {Salama}, {Savransky},
  {Schneider}, {Sivaramakrishnan}, {Song}, {Soummer}, {Thomas}, {Vasisht},
  {Wallace}, {Ward-Duong}, {Wiktorowicz}, {Wolff}, \&
  {Zuckerman}}]{Macintosh2015}
{Macintosh}, B., {Graham}, J.~R., {Barman}, T., {et~al.} 2015, Science, 350, 64

\bibitem[{{Madurowicz} {et~al.}(2019){Madurowicz}, {Macintosh}, {Chilcote},
  {Perrin}, {Poyneer}, {Pueyo}, {Ruffio}, {Bailey}, {Barman}, {Bulger},
  {Cotten}, {De Rosa}, {Doyon}, {Duch{\^e}ne}, {Esposito}, {Fitzgerald},
  {Follette}, {Gerard}, {Goodsell}, {Graham}, {Greenbaum}, {Hibon}, {Hung},
  {Ingraham}, {Kalas}, {Konopacky}, {Maire}, {Marchis}, {Marley}, {Marois},
  {Metchev}, {Millar-Blanchaer}, {Nielsen}, {Oppenheimer}, {Palmer},
  {Patience}, {Rajan}, {Rameau}, {Rantakyr{\"o}}, {Savransky},
  {Sivaramakrishnan}, {Song}, {Soummer}, {Tallis}, {Thomas}, {Wang},
  {Ward-Duong}, \& {Wolff}}]{Madurowicz2019}
{Madurowicz}, A., {Macintosh}, B., {Chilcote}, J., {et~al.} 2019, Journal of
  Astronomical Telescopes, Instruments, and Systems, 5, 049003

\bibitem[{{Maire} {et~al.}(2018){Maire}, {Rodet}, {Lazzoni}, {Boccaletti},
  {Brandner}, {Galicher}, {Cantalloube}, {Mesa}, {Klahr}, {Beust}, {Chauvin},
  {Desidera}, {Janson}, {Keppler}, {Olofsson}, {Augereau}, {Daemgen},
  {Henning}, {Th{\'e}bault}, {Bonnefoy}, {Feldt}, {Gratton}, {Lagrange},
  {Langlois}, {Meyer}, {Vigan}, {D'Orazi}, {Hagelberg}, {Le Coroller}, {Ligi},
  {Rouan}, {Samland}, {Schmidt}, {Udry}, {Zurlo}, {Abe}, {Carle},
  {Delboulb{\'e}}, {Feautrier}, {Magnard}, {Maurel}, {Moulin}, {Pavlov},
  {Perret}, {Petit}, {Ramos}, {Rigal}, {Roux}, \& {Weber}}]{Maire2018}
{Maire}, A.~L., {Rodet}, L., {Lazzoni}, C., {et~al.} 2018, \aap, 615, A177

\bibitem[{{Marois} {et~al.}(2008){Marois}, {Lafreni{\`e}re}, {Macintosh}, \&
  {Doyon}}]{Marois2008}
{Marois}, C., {Lafreni{\`e}re}, D., {Macintosh}, B., \& {Doyon}, R. 2008, \apj,
  673, 647

\bibitem[{{Marois} {et~al.}(2010){Marois}, {Zuckerman}, {Konopacky},
  {Macintosh}, \& {Barman}}]{Marois2010}
{Marois}, C., {Zuckerman}, B., {Konopacky}, Q.~M., {Macintosh}, B., \&
  {Barman}, T. 2010, \nat, 468, 1080

\bibitem[{{Martinez} {et~al.}(2013){Martinez}, {Kasper}, {Costille}, {Sauvage},
  {Dohlen}, {Puget}, \& {Beuzit}}]{Martinez2013}
{Martinez}, P., {Kasper}, M., {Costille}, A., {et~al.} 2013, \aap, 554, A41

\bibitem[{{Mawet} {et~al.}(2014){Mawet}, {Milli}, {Wahhaj}, {Pelat}, {Absil},
  {Delacroix}, {Boccaletti}, {Kasper}, {Kenworthy}, {Marois}, {Mennesson}, \&
  {Pueyo}}]{Mawet2014}
{Mawet}, D., {Milli}, J., {Wahhaj}, Z., {et~al.} 2014, \apj, 792, 97

\bibitem[{{Mesa} {et~al.}(2018){Mesa}, {Baudino}, {Charnay}, {D'Orazi},
  {Desidera}, {Boccaletti}, {Gratton}, {Bonnefoy}, {Delorme}, {Langlois},
  {Vigan}, {Zurlo}, {Maire}, {Janson}, {Antichi}, {Baruffolo}, {Bruno},
  {Cascone}, {Chauvin}, {Claudi}, {De Caprio}, {Fantinel}, {Farisato}, {Feldt},
  {Giro}, {Hagelberg}, {Incorvaia}, {Lagadec}, {Lagrange}, {Lazzoni}, {Lessio},
  {Salasnich}, {Scuderi}, {Sissa}, \& {Turatto}}]{Mesa2018}
{Mesa}, D., {Baudino}, J.~L., {Charnay}, B., {et~al.} 2018, \aap, 612, A92

\bibitem[{{Monnier} {et~al.}(2007){Monnier}, {Zhao}, {Pedretti}, {Thureau},
  {Ireland}, {Muirhead}, {Berger}, {Millan-Gabet}, {Van Belle}, {ten
  Brummelaar}, {McAlister}, {Ridgway}, {Turner}, {Sturmann}, {Sturmann}, \&
  {Berger}}]{Monnier2007}
{Monnier}, J.~D., {Zhao}, M., {Pedretti}, E., {et~al.} 2007, Science, 317, 342

\bibitem[{{Morzinski} {et~al.}(2014){Morzinski}, {Close}, {Males}, {Kopon},
  {Hinz}, {Esposito}, {Riccardi}, {Puglisi}, {Pinna}, {Briguglio}, {Xompero},
  {Quir{\'o}s-Pacheco}, {Bailey}, {Follette}, {Rodigas}, {Wu}, {Arcidiacono},
  {Argomedo}, {Busoni}, {Hare}, {Uomoto}, \& {Weinberger}}]{Morzinski2014}
{Morzinski}, K.~M., {Close}, L.~M., {Males}, J.~R., {et~al.} 2014, in Society
  of Photo-Optical Instrumentation Engineers (SPIE) Conference Series, Vol.
  9148, Adaptive Optics Systems IV, ed. E.~{Marchetti}, L.~M. {Close}, \& J.-P.
  {Vran}, 914804

\bibitem[{{Morzinski} {et~al.}(2015){Morzinski}, {Males}, {Skemer}, {Close},
  {Hinz}, {Rodigas}, {Puglisi}, {Esposito}, {Riccardi}, {Pinna}, {Xompero},
  {Briguglio}, {Bailey}, {Follette}, {Kopon}, {Weinberger}, \&
  {Wu}}]{Morzinski2015}
{Morzinski}, K.~M., {Males}, J.~R., {Skemer}, A.~J., {et~al.} 2015, \apj, 815,
  108

\bibitem[{Nelder \& Mead(1965)}]{Nelder1965}
Nelder, J.~A. \& Mead, R. 1965, The Computer Journal, 7, 308

\bibitem[{{Nielsen} {et~al.}(2019){Nielsen}, {De Rosa}, {Macintosh}, {Wang},
  {Ruffio}, {Chiang}, {Marley}, {Saumon}, {Savransky}, {Ammons}, {Bailey},
  {Barman}, {Blain}, {Bulger}, {Burrows}, {Chilcote}, {Cotten}, {Czekala},
  {Doyon}, {Duch{\^e}ne}, {Esposito}, {Fabrycky}, {Fitzgerald}, {Follette},
  {Fortney}, {Gerard}, {Goodsell}, {Graham}, {Greenbaum}, {Hibon}, {Hinkley},
  {Hirsch}, {Hom}, {Hung}, {Dawson}, {Ingraham}, {Kalas}, {Konopacky},
  {Larkin}, {Lee}, {Lin}, {Maire}, {Marchis}, {Marois}, {Metchev},
  {Millar-Blanchaer}, {Morzinski}, {Oppenheimer}, {Palmer}, {Patience},
  {Perrin}, {Poyneer}, {Pueyo}, {Rafikov}, {Rajan}, {Rameau}, {Rantakyr{\"o}},
  {Ren}, {Schneider}, {Sivaramakrishnan}, {Song}, {Soummer}, {Tallis},
  {Thomas}, {Ward-Duong}, \& {Wolff}}]{Nielsen2019}
{Nielsen}, E.~L., {De Rosa}, R.~J., {Macintosh}, B., {et~al.} 2019, \aj, 158,
  13

\bibitem[{{Nielsen} {et~al.}(2013){Nielsen}, {Liu}, {Wahhaj}, {Biller},
  {Hayward}, {Close}, {Males}, {Skemer}, {Chun}, {Ftaclas}, {Alencar},
  {Artymowicz}, {Boss}, {Clarke}, {de Gouveia Dal Pino}, {Gregorio-Hetem},
  {Hartung}, {Ida}, {Kuchner}, {Lin}, {Reid}, {Shkolnik}, {Tecza}, {Thatte}, \&
  {Toomey}}]{Nielsen2013}
{Nielsen}, E.~L., {Liu}, M.~C., {Wahhaj}, Z., {et~al.} 2013, \apj, 776, 4

\bibitem[{{Otten} {et~al.}(2017){Otten}, {Snik}, {Kenworthy}, {Keller},
  {Males}, {Morzinski}, {Close}, {Codona}, {Hinz}, {Hornburg}, {Brickson}, \&
  {Escuti}}]{Otten2017}
{Otten}, G. P.~P.~L., {Snik}, F., {Kenworthy}, M.~A., {et~al.} 2017, \apj, 834,
  175

\bibitem[{{Otten} {et~al.}(2014){Otten}, {Snik}, {Kenworthy}, {Miskiewicz}, \&
  {Escuti}}]{Otten2014}
{Otten}, G. P.~P.~L., {Snik}, F., {Kenworthy}, M.~A., {Miskiewicz}, M.~N., \&
  {Escuti}, M.~J. 2014, Optics Express, 22, 30287

\bibitem[{Pedregosa {et~al.}(2011)Pedregosa, Varoquaux, Gramfort, Michel,
  Thirion, Grisel, Blondel, Prettenhofer, Weiss, Dubourg, Vanderplas, Passos,
  Cournapeau, Brucher, Perrot, \& Duchesnay}]{Scikit-learn2011}
Pedregosa, F., Varoquaux, G., Gramfort, A., {et~al.} 2011, Journal of Machine
  Learning Research, 12, 2825

\bibitem[{{Rajan} {et~al.}(2017){Rajan}, {Rameau}, {De Rosa}, {Marley},
  {Graham}, {Macintosh}, {Marois}, {Morley}, {Patience}, {Pueyo}, {Saumon},
  {Ward-Duong}, {Ammons}, {Arriaga}, {Bailey}, {Barman}, {Bulger}, {Burrows},
  {Chilcote}, {Cotten}, {Czekala}, {Doyon}, {Duch{\^e}ne}, {Esposito},
  {Fitzgerald}, {Follette}, {Fortney}, {Goodsell}, {Greenbaum}, {Hibon},
  {Hung}, {Ingraham}, {Johnson-Groh}, {Kalas}, {Konopacky}, {Lafreni{\`e}re},
  {Larkin}, {Maire}, {Marchis}, {Metchev}, {Millar-Blanchaer}, {Morzinski},
  {Nielsen}, {Oppenheimer}, {Palmer}, {Patel}, {Perrin}, {Poyneer},
  {Rantakyr{\"o}}, {Ruffio}, {Savransky}, {Schneider}, {Sivaramakrishnan},
  {Song}, {Soummer}, {Thomas}, {Vasisht}, {Wallace}, {Wang}, {Wiktorowicz}, \&
  {Wolff}}]{Rajan2017}
{Rajan}, A., {Rameau}, J., {De Rosa}, R.~J., {et~al.} 2017, \aj, 154, 10

\bibitem[{{Ruffio} {et~al.}(2017){Ruffio}, {Macintosh}, {Wang}, {Pueyo},
  {Nielsen}, {De Rosa}, {Czekala}, {Marley}, {Arriaga}, {Bailey}, {Barman},
  {Bulger}, {Chilcote}, {Cotten}, {Doyon}, {Duch{\^e}ne}, {Fitzgerald},
  {Follette}, {Gerard}, {Goodsell}, {Graham}, {Greenbaum}, {Hibon}, {Hung},
  {Ingraham}, {Kalas}, {Konopacky}, {Larkin}, {Maire}, {Marchis}, {Marois},
  {Metchev}, {Millar-Blanchaer}, {Morzinski}, {Oppenheimer}, {Palmer},
  {Patience}, {Perrin}, {Poyneer}, {Rajan}, {Rameau}, {Rantakyr{\"o}},
  {Savransky}, {Schneider}, {Sivaramakrishnan}, {Song}, {Soummer}, {Thomas},
  {Wallace}, {Ward-Duong}, {Wiktorowicz}, \& {Wolff}}]{Ruffio2017}
{Ruffio}, J.-B., {Macintosh}, B., {Wang}, J.~J., {et~al.} 2017, \apj, 842, 14

\bibitem[{{Ruffio} {et~al.}(2018){Ruffio}, {Mawet}, {Czekala}, {Macintosh}, {De
  Rosa}, {Ruane}, {Bottom}, {Pueyo}, {Wang}, {Hirsch}, {Zhu}, \&
  {Nielsen}}]{Ruffio2018}
{Ruffio}, J.-B., {Mawet}, D., {Czekala}, I., {et~al.} 2018, \aj, 156, 196

\bibitem[{{Samland} {et~al.}(2021){Samland}, {Bouwman}, {Hogg}, {Brandner},
  {Henning}, \& {Janson}}]{Samland2021}
{Samland}, M., {Bouwman}, J., {Hogg}, D.~W., {et~al.} 2021, \aap, 646, A24

\bibitem[{{Samland} {et~al.}(2017){Samland}, {Molli{\`e}re}, {Bonnefoy},
  {Maire}, {Cantalloube}, {Cheetham}, {Mesa}, {Gratton}, {Biller}, {Wahhaj},
  {Bouwman}, {Brandner}, {Melnick}, {Carson}, {Janson}, {Henning}, {Homeier},
  {Mordasini}, {Langlois}, {Quanz}, {van Boekel}, {Zurlo}, {Schlieder},
  {Avenhaus}, {Beuzit}, {Boccaletti}, {Bonavita}, {Chauvin}, {Claudi}, {Cudel},
  {Desidera}, {Feldt}, {Fusco}, {Galicher}, {Kopytova}, {Lagrange}, {Le
  Coroller}, {Martinez}, {Moeller-Nilsson}, {Mouillet}, {Mugnier}, {Perrot},
  {Sevin}, {Sissa}, {Vigan}, \& {Weber}}]{Samland2017}
{Samland}, M., {Molli{\`e}re}, P., {Bonnefoy}, M., {et~al.} 2017, \aap, 603,
  A57

\bibitem[{{Sivanandam} {et~al.}(2006){Sivanandam}, {Hinz}, {Heinze}, {Freed},
  \& {Breuninger}}]{Sivanandam2006}
{Sivanandam}, S., {Hinz}, P.~M., {Heinze}, A.~N., {Freed}, M., \& {Breuninger},
  A.~H. 2006, in Society of Photo-Optical Instrumentation Engineers (SPIE)
  Conference Series, Vol. 6269, Society of Photo-Optical Instrumentation
  Engineers (SPIE) Conference Series, ed. I.~S. {McLean} \& M.~{Iye}, 62690U

\bibitem[{{Skemer} {et~al.}(2015){Skemer}, {Hinz}, {Montoya}, {Skrutskie},
  {Leisenring}, {Durney}, {Woodward}, {Wilson}, {Nelson}, {Bailey}, {Defrere},
  \& {Stone}}]{Skemer2015}
{Skemer}, A.~J., {Hinz}, P., {Montoya}, M., {et~al.} 2015, in Society of
  Photo-Optical Instrumentation Engineers (SPIE) Conference Series, Vol. 9605,
  Techniques and Instrumentation for Detection of Exoplanets VII, ed.
  S.~{Shaklan}, 96051D

\bibitem[{{Skemer} {et~al.}(2014){Skemer}, {Marley}, {Hinz}, {Morzinski},
  {Skrutskie}, {Leisenring}, {Close}, {Saumon}, {Bailey}, {Briguglio},
  {Defrere}, {Esposito}, {Follette}, {Hill}, {Males}, {Puglisi}, {Rodigas}, \&
  {Xompero}}]{Skemer2014}
{Skemer}, A.~J., {Marley}, M.~S., {Hinz}, P.~M., {et~al.} 2014, \apj, 792, 17

\bibitem[{{Skrutskie} {et~al.}(2010){Skrutskie}, {Jones}, {Hinz}, {Garnavich},
  {Wilson}, {Nelson}, {Solheid}, {Durney}, {Hoffmann}, {Vaitheeswaran},
  {McMahon}, {Leisenring}, \& {Wong}}]{Skrutskie2010}
{Skrutskie}, M.~F., {Jones}, T., {Hinz}, P., {et~al.} 2010, in Society of
  Photo-Optical Instrumentation Engineers (SPIE) Conference Series, Vol. 7735,
  Ground-based and Airborne Instrumentation for Astronomy III, ed. I.~S.
  {McLean}, S.~K. {Ramsay}, \& H.~{Takami}, 77353H

\bibitem[{{Snik} {et~al.}(2012){Snik}, {Otten}, {Kenworthy}, {Miskiewicz},
  {Escuti}, {Packham}, \& {Codona}}]{Snik2012}
{Snik}, F., {Otten}, G., {Kenworthy}, M., {et~al.} 2012, in Society of
  Photo-Optical Instrumentation Engineers (SPIE) Conference Series, Vol. 8450,
  Modern Technologies in Space- and Ground-based Telescopes and Instrumentation
  II, ed. R.~{Navarro}, C.~R. {Cunningham}, \& E.~{Prieto}, 84500M

\bibitem[{{Soummer} {et~al.}(2012){Soummer}, {Pueyo}, \&
  {Larkin}}]{Soummer2012}
{Soummer}, R., {Pueyo}, L., \& {Larkin}, J. 2012, \apjl, 755, L28

\bibitem[{{Stolker} {et~al.}(2019){Stolker}, {Bonse}, {Quanz}, {Amara},
  {Cugno}, {Bohn}, \& {Boehle}}]{Stolker2019}
{Stolker}, T., {Bonse}, M.~J., {Quanz}, S.~P., {et~al.} 2019, \aap, 621, A59

\bibitem[{{Stone} {et~al.}(2022){Stone}, {Skemer}, {Hinz}, {Ertel},
  {Briesemeister}, {Leisenring}, {Durney}, {Montoya}, {Woodward}, {Skrutskie},
  \& {Barman}}]{Stone2022}
{Stone}, J.~M., {Skemer}, A., {Hinz}, P., {et~al.} 2022, in Society of
  Photo-Optical Instrumentation Engineers (SPIE) Conference Series, Vol. 12184,
  Ground-based and Airborne Instrumentation for Astronomy IX, ed. C.~J.
  {Evans}, J.~J. {Bryant}, \& K.~{Motohara}, 1218442

\bibitem[{{Sutlieff} {et~al.}(2023){Sutlieff}, {Birkby}, {Stone}, {Doelman},
  {Kenworthy}, {Panwar}, {Bohn}, {Ertel}, {Snik}, {Woodward}, {Skemer},
  {Leisenring}, {Strassmeier}, \& {Charbonneau}}]{Sutlieff2023}
{Sutlieff}, B.~J., {Birkby}, J.~L., {Stone}, J.~M., {et~al.} 2023, \mnras, 520,
  4235

\bibitem[{{Sutlieff} {et~al.}(2021){Sutlieff}, {Bohn}, {Birkby}, {Kenworthy},
  {Morzinski}, {Doelman}, {Males}, {Snik}, {Close}, {Hinz}, \&
  {Charbonneau}}]{Sutlieff2021}
{Sutlieff}, B.~J., {Bohn}, A.~J., {Birkby}, J.~L., {et~al.} 2021, \mnras, 506,
  3224

\bibitem[{{Vigan} {et~al.}(2021){Vigan}, {Fontanive}, {Meyer}, {Biller},
  {Bonavita}, {Feldt}, {Desidera}, {Marleau}, {Emsenhuber}, {Galicher}, {Rice},
  {Forgan}, {Mordasini}, {Gratton}, {Le Coroller}, {Maire}, {Cantalloube},
  {Chauvin}, {Cheetham}, {Hagelberg}, {Lagrange}, {Langlois}, {Bonnefoy},
  {Beuzit}, {Boccaletti}, {D'Orazi}, {Delorme}, {Dominik}, {Henning}, {Janson},
  {Lagadec}, {Lazzoni}, {Ligi}, {Menard}, {Mesa}, {Messina}, {Moutou},
  {M{\"u}ller}, {Perrot}, {Samland}, {Schmid}, {Schmidt}, {Sissa}, {Turatto},
  {Udry}, {Zurlo}, {Abe}, {Antichi}, {Asensio-Torres}, {Baruffolo}, {Baudoz},
  {Baudrand}, {Bazzon}, {Blanchard}, {Bohn}, {Brown Sevilla}, {Carbillet},
  {Carle}, {Cascone}, {Charton}, {Claudi}, {Costille}, {De Caprio},
  {Delboulb{\'e}}, {Dohlen}, {Engler}, {Fantinel}, {Feautrier}, {Fusco},
  {Gigan}, {Girard}, {Giro}, {Gisler}, {Gluck}, {Gry}, {Hubin}, {Hugot},
  {Jaquet}, {Kasper}, {Le Mignant}, {Llored}, {Madec}, {Magnard}, {Martinez},
  {Maurel}, {M{\"o}ller-Nilsson}, {Mouillet}, {Moulin}, {Orign{\'e}}, {Pavlov},
  {Perret}, {Petit}, {Pragt}, {Puget}, {Rabou}, {Ramos}, {Rickman}, {Rigal},
  {Rochat}, {Roelfsema}, {Rousset}, {Roux}, {Salasnich}, {Sauvage}, {Sevin},
  {Soenke}, {Stadler}, {Suarez}, {Wahhaj}, {Weber}, \& {Wildi}}]{Vigan2021}
{Vigan}, A., {Fontanive}, C., {Meyer}, M., {et~al.} 2021, \aap, 651, A72

\bibitem[{{Virtanen} {et~al.}(2020){Virtanen}, {Gommers}, {Oliphant},
  {Haberland}, {Reddy}, {Cournapeau}, {Burovski}, {Peterson}, {Weckesser},
  {Bright}, {van der Walt}, {Brett}, {Wilson}, {Jarrod Millman}, {Mayorov},
  {Nelson}, {Jones}, {Kern}, {Larson}, {Carey}, {Polat}, {Feng}, {Moore}, {Vand
  erPlas}, {Laxalde}, {Perktold}, {Cimrman}, {Henriksen}, {Quintero}, {Harris},
  {Archibald}, {Ribeiro}, {Pedregosa}, {van Mulbregt}, \&
  {Contributors}}]{Scipy2020}
{Virtanen}, P., {Gommers}, R., {Oliphant}, T.~E., {et~al.} 2020, Nature
  Methods, 17, 261

\bibitem[{{Wagner} {et~al.}(2020){Wagner}, {Stone}, {Dong}, {Ertel}, {Apai},
  {Doelman}, {Bohn}, {Najita}, {Brittain}, {Kenworthy}, {Keppler}, {Webster},
  {Mailhot}, \& {Snik}}]{Wagner2020}
{Wagner}, K., {Stone}, J., {Dong}, R., {et~al.} 2020, \aj, 159, 252

\bibitem[{{Wahhaj} {et~al.}(2015){Wahhaj}, {Cieza}, {Mawet}, {Yang}, {Canovas},
  {de Boer}, {Casassus}, {M{\'e}nard}, {Schreiber}, {Liu}, {Biller}, {Nielsen},
  \& {Hayward}}]{Wahhaj2015}
{Wahhaj}, Z., {Cieza}, L.~A., {Mawet}, D., {et~al.} 2015, \aap, 581, A24

\bibitem[{{Wang} {et~al.}(2018){Wang}, {Graham}, {Dawson}, {Fabrycky}, {De
  Rosa}, {Pueyo}, {Konopacky}, {Macintosh}, {Marois}, {Chiang}, {Ammons},
  {Arriaga}, {Bailey}, {Barman}, {Bulger}, {Chilcote}, {Cotten}, {Doyon},
  {Duch{\^e}ne}, {Esposito}, {Fitzgerald}, {Follette}, {Gerard}, {Goodsell},
  {Greenbaum}, {Hibon}, {Hung}, {Ingraham}, {Kalas}, {Larkin}, {Maire},
  {Marchis}, {Marley}, {Metchev}, {Millar-Blanchaer}, {Nielsen}, {Oppenheimer},
  {Palmer}, {Patience}, {Perrin}, {Poyneer}, {Rajan}, {Rameau},
  {Rantakyr{\"o}}, {Ruffio}, {Savransky}, {Schneider}, {Sivaramakrishnan},
  {Song}, {Soummer}, {Thomas}, {Wallace}, {Ward-Duong}, {Wiktorowicz}, \&
  {Wolff}}]{Wang2018}
{Wang}, J.~J., {Graham}, J.~R., {Dawson}, R., {et~al.} 2018, \aj, 156, 192

\bibitem[{{Zurlo} {et~al.}(2016){Zurlo}, {Vigan}, {Galicher}, {Maire}, {Mesa},
  {Gratton}, {Chauvin}, {Kasper}, {Moutou}, {Bonnefoy}, {Desidera}, {Abe},
  {Apai}, {Baruffolo}, {Baudoz}, {Baudrand}, {Beuzit}, {Blancard},
  {Boccaletti}, {Cantalloube}, {Carle}, {Cascone}, {Charton}, {Claudi},
  {Costille}, {de Caprio}, {Dohlen}, {Dominik}, {Fantinel}, {Feautrier},
  {Feldt}, {Fusco}, {Gigan}, {Girard}, {Gisler}, {Gluck}, {Gry}, {Henning},
  {Hugot}, {Janson}, {Jaquet}, {Lagrange}, {Langlois}, {Llored}, {Madec},
  {Magnard}, {Martinez}, {Maurel}, {Mawet}, {Meyer}, {Milli},
  {Moeller-Nilsson}, {Mouillet}, {Orign{\'e}}, {Pavlov}, {Petit}, {Puget},
  {Quanz}, {Rabou}, {Ramos}, {Rousset}, {Roux}, {Salasnich}, {Salter},
  {Sauvage}, {Schmid}, {Soenke}, {Stadler}, {Suarez}, {Turatto}, {Udry},
  {Vakili}, {Wahhaj}, {Wildi}, \& {Antichi}}]{Zurlo2016}
{Zurlo}, A., {Vigan}, A., {Galicher}, R., {et~al.} 2016, \aap, 587, A57

\end{thebibliography}

\begin{appendix}
\section{Reduction of the Altair dataset}
\label{apd}
The Altair dataset shares similar properties as the HR~2562 dataset: significant linear PSF drifts between frames and similar parallactic angle range. We used the same data reduction method of the HR~2562 dataset to pre-process and post-process the Altair dataset as described in Sect.~\ref{sec:algorithms on gvAPP180}. Because the central star Altair was saturated, we used cross-correlation to align the images. The star PSF model was created by scaling up the calibration PSF to match the integration time of the science frames. The frames were binned every 20 frames. A master sky background made by the median of the 20 sky frames was subtracted from the science frames.

To verify the result found for HR~2562 that choosing reference pixels from the dark hole of the same PSF is better than choosing reference pixels from other regions, we did five injections at a separation of 5\,$\lambda/D$, the middle of the upper dark hole. Each time we injected a fake planet at a different angle and reduced the upper dark region using the four reference pixel choices: only from the upper dark hole, only from the upper bright side, only from the lower dark hole, and only from the lower bright side. The injected signal was the 5$\sigma$ contrast of that position reduced by choosing reference pixels from the upper dark hole. We calculated the retrieved S/N, contrast deviation and position deviation to the injected values. We averaged the results of the five injections for each reference pixel choice and took the standard deviation as the 1$\sigma$ error as presented in Table~\ref{Table:Altair_injection}.

\end{appendix}

\end{document}